\documentclass[12pt,preprint]{aastex}

\def\ltsima{$\; \buildrel < \over \sim \;$}
\def\lsim{\lower.5ex\hbox{\ltsima}}
\def\gtsima{$\; \buildrel > \over \sim \;$}
\def\gsim{\lower.5ex\hbox{\gtsima}}

\shorttitle{Evolution of Galaxies at Rest-frame 1500\AA~and 2800\AA}
\shortauthors{Dahlen et al.}

\begin{document}
\title{Evolution of the Luminosity Function, Star Formation Rate, Morphology and Size of Star-forming Galaxies Selected at Rest-frame 1500\AA~and 2800\AA}


\author{
Tomas Dahlen\altaffilmark{1,2}, 
Bahram Mobasher\altaffilmark{2,3},
Mark Dickinson\altaffilmark{4}, 
Henry C. Ferguson\altaffilmark{2},
Mauro Giavalisco\altaffilmark{2},
Claudia Kretchmer,\altaffilmark{5},
and Swara Ravindranath\altaffilmark{6}
}
\email{dahlen@physto.se}

\altaffiltext{1}{Department of Physics, Stockholm University, SE-10961, Stockholm, Sweden; dahlen@physto.se}
\altaffiltext{2}{Space Telescope Science Institute, Baltimore, MD 21218; mobasher@stsci.edu, ferguson@stsci.edu, mauro@stsci.edu}
\altaffiltext{3}{Affiliated with the Space Telescope Division of the European Space Agency,
ESTEC, Noordwijk, Netherlands.}
\altaffiltext{4}{National Optical Astronomy Observatory, P.O. Box 26732, Tucson, AZ 85726; med@noao.edu.}
\altaffiltext{5}{Department of Physics and Astronomy, Johns Hopkins University, Baltimore, MD 21218; claudia@pha.jhu.edu}
\altaffiltext{6}{Inter-University Center for Astronomy \& Astrophysics, Post Bag - 4, Ganeshkhind, Pune, India 411007; swara@iucaa.ernet.in}


\begin{abstract}
Using the multiwavelength photometric and spectroscopic data 
covering the Chandra Deep Field South obtained 
within the Great Observatories Origins Deep Survey, we investigate the 
rest-frame UV properties of galaxies to $z\sim 2.2$,
including the evolution of the luminosity function, the luminosity density,
star formation rate (SFR) and galaxy morphology. We find a significant 
brightening 
($\sim 1$~mag) in the rest-frame 2800\AA~characteristic magnitude
($M^\ast$) over the redshift range $0.3\lsim z\lsim1.7$~and no evolution at
higher redshifts. The rest-frame 2800\AA~luminosity
density shows an increase by a factor $\sim 4$~over the redshift range 
investigated. 
We estimate the star formation rate density to $z\sim$2.2 
from the 1500\AA~and 2800\AA~luminosities. When no correction for
extinction is made, we find that the star formation
rate derived from the 2800\AA~luminosity density is almost factor 
two higher than that derived from the 1500\AA~luminosities. 
Attributing this difference to
differential dust extinction, we find that an E(B--V)=0.20 results
in the same extinction corrected star formation rate from both
1500\AA~and 2800\AA~luminosities. The extinction corrected SFR is a 
factor $\sim 6.5$~($\sim 3.7$) higher than the uncorrected SFR derived 
from 1500\AA~(2800\AA) luminosity. 
We investigate the morphological
composition of our sample by fitting S\'{e}rsic profiles to the 
$HST$~ACS galaxy images at a fixed rest-frame wavelength of 2800\AA~at 
$0.5\lsim z\lsim2.2$. 
We find that the fraction of apparently bulge-dominated galaxies
(S\'{e}rsic index $n>2.5$) increases from $\sim$10\%~at $z\sim 0.5$~to 
$\sim$30\%~at $z\sim 2.2$.
At the same time, we note that galaxies get bluer at increasing redshift.
This suggests a scenario where an increased fraction of the star formation
takes place in bulge-dominated systems at high redshift. 
This could be the evidence that the present day ellipticals are a result of
assembly (i.e., mergers) of galaxies at $z\gsim$1.
Finally, we
find that galaxy sizes for a luminosity selected sample evolves
as $r_h\propto (1+z)^{-1.1}$~between redshifts $z=2.2$~and $z=1.1$.
This is consistent with previous measurements and suggests a 
similar evolution over the redshift range $0\lsim z \lsim$6.
\end{abstract}


\section{Introduction}
Over the last decade, a large number of studies have measured the evolution of
cosmic star formation rate (SFR) in the range $0 < z < 6$, using different
diagnostics (Steidel et al. 1995; Madau et al. 1996; Chary \& Elbaz 2001; 
Giavalisco et al. 2004a; Schiminovich et al. 2005; Takeuchi et al. 2005; see also
Hopkins (2004) for a compilation of the SFR from a large number of 
multiwavelength surveys). These studies show a rapid increase in the SFR 
to $z\sim 1-2$, beyond which it flattens or turns over. While most studies are 
in broad agreement on the shape of the SFR, the absolute normalization is still
uncertain, mainly due to unknown extinction in the UV to optical bands and 
systematic effects depending on star formation diagnostic, selection of
the star-forming population and the adopted flux limit of the surveys used 
(e.g., Hopkins \& Beacom 2006). A further question in these investigations 
concerns the morphology of star forming galaxies and how it evolves with redshift. 
These studies require complete surveys of star forming galaxies with known 
redshifts, SFRs and morphologies. 

Using the high spatial resolution and sensitivity of the Advance Camera
for Surveys (ACS) on-board the Hubble Space Telescope (HST), a number of
large, multi-waveband and deep surveys have now been completed, with the
two deepest being the Great Observatories Origins Deep Survey (GOODS; 
Giavalisco et al. 2004b) and the Hubble Ultra-Deep Field (HUDF; Beckwith, 
S. V. W., et al. 2006, in preparation). The combined wavelength coverage, 
spatial resolution and
depth of these surveys allow accurate measurements of the SFR, morphology
and size of these galaxies to faint flux levels ($m_z\gsim 25$) and high
redshifts ($z\sim 6$). 

Recently, the ultraviolet morphological properties of star forming galaxies 
in the GOODS
and HUDF were explored by a number of investigators. Using WFPC2
observations of the GOODS parallel fields at F300W, 
de Mello et al. (2006) studied rest-frame UV properties of 
galaxies to $z\sim 1.5$. They found that their sample included all
major morphological types, with compact and peculiar morphologies 
becoming relatively more abundant at higher redshift ($z\gsim 0.7$).
Furthermore, Lotz et al. (2006) studied the UV morphologies of star-forming 
galaxies in the GOODS-S at $z\sim 1.5$~and $z\sim 4$ and found no significant 
differences between the galaxy morphologies at these redshifts. 
Conselice et al. (2004) used the $HST$~ACS observations of GOODS-S
to identify populations of luminous diffuse objects and
luminous asymmetric objects at $1<z<2$~which they argue are the  
progenitors of todays normal disk and elliptical galaxies.

Ravindranath et al. (2006) find that among their sample
of bright LBGs at $z>2.5$, about 40\% have light profiles that can be
approximated by an exponential profile as seen in disks, while 30\%
have close to r$^{1/4}$ profiles as seen spheroids. However, they note
that even these galaxies do show clumpy or faint asymmetric features
characterestic of tidal interaction or minor merger. The method of
Gini coefficients used by Lotz et al. (2006) is more sensitive to
merger-like features. Lotz et al. (2006) find that among their
sample of LBGs at $z=4$, 30\% have relatively undisturbed spheroid-like
morphologies, about 10-25\% are major mergers. About 50\% are like
exponential disks or have minor mergers. Except for small differences
arising from the different methods that are employed, the results
from Ravindranath et al. (2006) are in good agreement with Lotz
et al. (2006).
   
Finally, the evolution of galaxy sizes has also been investigated using 
the GOODS and HUDF data sets by Bouwens et al. (2004a, 2004b, 2005) and 
Ferguson et al. (2004). 
In this paper, we study UV properties of a complete sample of
star-forming galaxies in the GOODS-S at rest-frame 1500\AA~and 2800\AA~ 
wavelengths. The investigation includes a study of the evolution of the 
luminosity function, the luminosity density, the SFR, morphology, and 
galaxy size. In \S 2 we describe our data. 
This is followed by results on the luminosity functions and luminosity 
densities (\S 3), the SFR (\S 4), and morphology and size (\S 5). 
We conclude and summarize our results in \S 6.

Throughout this paper we use $\Omega_{\Lambda}=0.7$, $\Omega_M=0.3$~
and $H_0=70$ km s$^{-1}$ Mpc$^{-1}$. Magnitudes are in the AB system.

\section{The GOODS data}
The GOODS South data used here consist of 
deep wide-field $HST$~ACS observations in the F435W, F606W, F775W, and F850LP
passbands (hereafter $B-,~V-,~i-$~and $z-$bands) combined with
ground-based ESO VLT/ISAAC near-IR $J-,~H-$, and $K_s-$band observations.
The limiting 10$\sigma$~sensitivity for the ACS observations are
$B=27.8$, $V=27.8$, $i=27.1$, and $z=26.6$~(Giavalisco et al. 2004b),
while the ISAAC observations have limits $J=24.8$, $H=24.2$,
and $K_s=24.1$~(Vandame, B., et al. 2006, in preparation).
In this investigation we use a $z$-band selected catalog and adopt a limiting 
magnitude $z<25.5$. We choose a 
limit $\sim 1$~mag brighter than the 10$\sigma$~sensitivity to assure
that we can reliably determine photometric redshifts and measure morphological
parameters for our galaxy sample. 
The area of the field used here is covered by both ACS and ISAAC
and corresponds to $\sim$130 square arcmin. 
Photometry is derived using SExtractor 
(Bertin \& Arnouts 1996). Before deriving photometry, we convolve the
space- and ground-based images
to a common psf. However, when deriving morphological properties, we
use original 'unconvolved' ACS images.

Besides the ACS+ISAAC photometric catalog, we also produce a solely 
ground-based catalog including observations in $U$-band (CTIO, 4 m MOSAIC),
$B-$, $V-$, $R-$~and $I-$bands (ESO, 2.2 m WFI), $J-,~H-$, and $Ks-$band
(ESO, NTT SOFI) and the above described ISAAC data. These observations are 
also centered on the GOODS South area, but the field is significantly wider, 
in total covering $\sim 1100$~square arcmin, of which about one third is 
observed in near-IR.
The 10$\sigma$~sensitivities are $U=25.9$, $B=26.2$, $V=25.8$, 
$R=25.8$, $i=23.5$, $J=22.8$, $H=22.0$, $K_s=21.8$ (Giavalisco et al. 2004b).
The limiting magnitude adopted for this $R$-band selected catalog is $R<25.0$.

Note that we restrict our investigation to GOODS South and do not include
GOODS North due to the shallower near-IR data of the latter, leading to
less accurate photometric redshifts at $z\gsim 1.5$~and faint magnitudes.
\subsection{Photometric redshifts}
Photometric redshifts are calculated using the template fitting
method incorporating priors as described in Dahlen et al. (2005).
For each object, we derive the photometric redshift, the redshift probability
distribution and the best-fitting spectral type. The spectral types cover
E, Sbc, Scd and Im templates (Coleman et al. 1980, extended in the UV and near-IR 
by Bolzonella et al. 2000) and two starburst templates (Kinney et al. 1996).

To test the accuracy of the photometric redshifts, we compare with a sample of
519 spectroscopic redshifts taken from the 
ESO/GOODS-CDFS spectroscopy master 
catalogue\footnote{http://www.eso.org/science/goods/spectroscopy/CDFS\_Mastercat/},
which is a compilation of a number of datasets with major contributions from
Le F\`{e}vre et al. (2004), Vanzetta et al. (2005) and Mignoli et al. (2005).
We find an accuracy $\Delta_z$=0.08 (where 
$\Delta_z\equiv\langle|z_{\rm phot}-z_{\rm spec}|/(1+z_{\rm spec})\rangle$), 
after excluding a small fraction ($\sim$3\%) outliers with $\Delta_z>0.3$.
Restricting the redshift range to that studied here, $0.29<z<2.37$, results in
a slightly lower outlier fraction ($\sim$2\%). Furthermore, the outliers have
distributions in redshift and magnitudes similar to the full redshift sample.
The distribution in spectral types somewhat biased towards later types for
the outliers, but there are outliers of all SED types. Therefore, we do expect 
to have a small fraction of outliers and galaxies with uncertain photometric 
redshifts. However, when deriving e.g., luminosity functions (hereafter LFs) and 
luminosity densities (LDs), we use a method that incorporates
the full photometric redshift probability distribution, instead of a single 
redshift, in order to minimize the effect of photometric redshift uncertainty.
This method is further described in Dahlen et al. (2005).

For a number of objects (312 of total 2976), we replace the photometric 
redshifts with available spectroscopic redshifts.
The spectral types of these objects are calculated using the photometric
redshift technique after fixing the redshift to its spectroscopic value. 

\subsection{Rest-frame 2800\AA}
One aim of this paper is to investigate the galaxy properties
at rest-frame 2800\AA. To do this, for each ACS band 
we identify a redshift 
range where the filter encompass the rest-frame 2800\AA~band.
This is defined so that at least 25\% of the filter's integrated 
transmission is short-ward of rest-frame 2800\AA~(lower redshift limit) and at 
least 25\% is long-ward (upper redshift limit).
Resulting redshift ranges are given in Table 1. The 
redshifts that divide the volume of each bin into two equal halves are 
$z=0.55$, $z=1.14$, $z=1.75$, and $z=2.23$~for
$B$-, $V$-, $i$- and $z$-bands, respectively. 
In the following when we derive galaxy morphologies, we use
 measurements in different bands depending on redshift
so that we always are probing the same rest-frame wavelength.  
By requiring that we observe the same rest-frame wavelength in all bins,
we minimize the effects of photometric K-corrections when deriving
rest-frame 2800\AA~luminosities, as well as morphological K-corrections, 
which may otherwise bias morphological interpretations (e.g., 
Papovich et al. 2005).

When deriving the LF, we also
use the ground-based catalog in order to measure rest-frame
2800\AA~at lower redshifts than possible with the ACS bands. 
Using the observed $U$-band, we probe the rest-frame 2800\AA~at 
$z=0.33$.
\subsection{Rest-frame 1500\AA}
To investigate how the UV derived SFR depends on the choice of
rest-frame band, we also calculate the LF at rest-frame 1500\AA~
using the same galaxies as in the 2800\AA~selected samples.
In the two lowest redshift bins ($z\sim 0.33$~and $z\sim 0.55$), 
we do not have any filter that observe sufficiently
close to the rest-frame 1500\AA~and do therefore not include these
bins. At $z\sim1.14$, we use the $B$-band, which 
observes rest-frame $\sim$2000\AA, and extrapolate to 1500\AA~using 
the best-fitting spectral template derived from the photometric
redshift fitting. In the two highest redshift bins, $z\sim1.75$, 
and $z\sim2.23$, we can more directly derive the rest-frame 1500\AA~
luminosity.
\section{The UV luminosity functions and luminosity densities}
\subsection{The 2800\AA~and 1500\AA~LFs}
We use the $1/V_{max}$-method (Schmidt 1968) to derive the LF. To account for 
the relatively large errors in the photometric redshifts, we incorporate the 
redshift probability distribution derived from the photometric redshift 
method when determining the LF. This procedure is described in detail in 
Dahlen et al. (2005). After deriving the LF, we fit the usual Schechter 
function (Schechter 1976) to the data.

At 2800\AA, we determine all the Schechter function parameters 
($M^*_{2800}$, $\alpha$~and $\phi_*$) in the redshift bin
at ($z\sim 0.55$), 
which is the bin where we reach the faintest limit in rest-frame 
2800\AA~absolute magnitude. In the remaining bins, we fix the faint-end 
slope to the value derived in this bin, $\alpha =-1.39$. The resulting Schechter 
function parameters are listed in Table 1.

\begin{figure}
\plotone{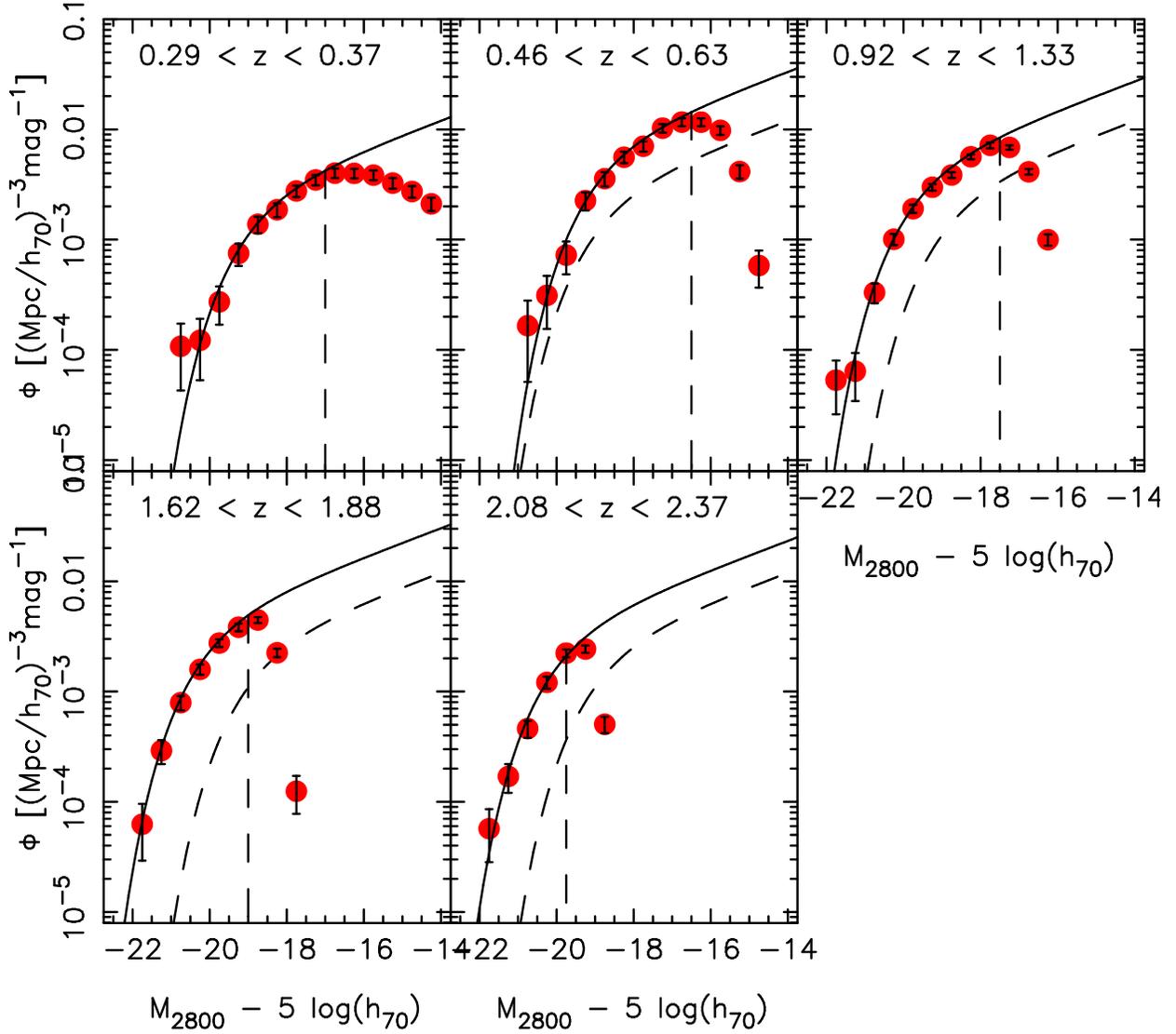}
\caption{The rest-frame 2800\AA~luminosity function. Best fitting Schechter 
functions are shown with solid lines. Completeness limit in each bin is shown 
with vertical dashed lines. We only use magnitude bins brighter than these 
limits when fitting Schechter functions. In each bin, we also show the best 
fitting Schechter function derived in the lowest redshift bin (dashed line). 
This illustrates the strong evolution in the 2800\AA~LF with redshift.
\label{fig1}}
\end{figure}

In Figure \ref{fig1}, we show the rest-frame 2800\AA~LFs in the 
five redshift bins. The best fitting Schechter functions are shown with solid 
lines, while the result derived in the lowest redshift bin is shown with 
dotted lines. Vertical dotted lines represent completeness limits. 
(When fitting the Schechter function parameters, we only use points 
brighter that these limits.) The completeness limits are derived by 
calculating the absolute 2800\AA~magnitude corresponding to the apparent
magnitude limit ($z=25.5$) for a range of galaxy SEDs from ellipticals
to starbursts at the central redshift of each bin. We thereafter choose 
the brightest of these magnitudes as the limit, ensuring us that we are 
complete for all considered galaxy types.

The figure clearly shows a strong evolution, 
where the characteristic magnitude gets brighter at higher redshifts.
The total brightening in $M_{2800}^*$~is $\sim 1$~mag between 
$z\sim 0.3$~and $z\sim 1.7$. This is similar to the evolution in
$M_{2800}^*$ found by Wolf et al. (2003) and Gabasch et al. (2004).
Between the two highest redshift bins, 
the evolution is consistent with being flat or slightly fading. 

\begin{figure}
\epsscale{0.8}
\plotone{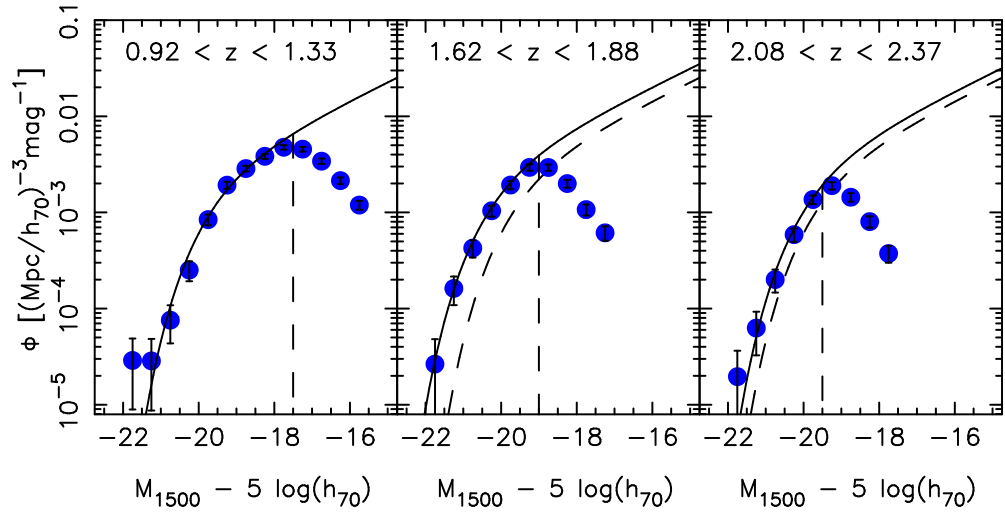}
\caption{The rest-frame 1500\AA~luminosity function. Best fitting Schechter 
functions are shown with solid lines. Completeness limit in each bin is shown 
with vertical dashed lines. We only use magnitude bins brighter than these 
limits when fitting Schechter functions. In each bin, we also show the best 
fitting Schechter function derived in the lowest redshift bin (dashed line). 
\label{fig2}}
\end{figure}

For the 1500\AA~LFs, we use the redshift bin at $z\sim 1.7$~to determine
the faint-end slope and find $\alpha=-1.48$. We chose this bin since it 
directly observes the rest-frame 1500\AA~(in contrast to the $z\sim 1.1$ 
bin which requires extrapolation). After fixing the faint-end slope, we 
derive the Schechter parameters. Results are given in Table 1 and 
Figure \ref{fig2}. 

Similar to the 2800\AA~results, the characteristic magnitude at 
1500\AA~shows a brightening with redshift from $z\sim$1.1 to $z\sim$1.7, 
with no significant evolution at higher redshift. 
Combined with results of Bouwens et al. (2005), who found 
that $M^*_{1350}$~brightens by $\sim 0.7$~mag between $z\sim 6$~and 
$z\sim 3$, we develop a scenario where
the characteristic UV magnitude brightens, turns over, and finally fades 
when going from high to low redshift. Such evolution is consistent with the 
hierarchical model (e.g., Cole et al. 2004) where the initial brightening at 
rest-frame UV is due to infall of gas,
mergers and merger induced star formation. When the merger rate, and the 
related star formation, decreases, galaxies start to fade as the reservoirs
of cold gas are depleted. The observations suggest 
a peak in the UV
characteristic magnitude at $z\sim 2-3$, which we note coincides with the 
proposed peak in the cosmic SFR, as well as the peak of the merger epoch for 
the brightest and most massive galaxies (Conselice et al. 2003). 
In contrast, we note that Gabasch et al. (2004) find a monotonic brightening
in the characteristic magnitude at both rest-frame 1500\AA~and 
2800\AA~to redshift $z\sim4-5$.

\subsection{The UV luminosity densities}
\begin{figure}
\epsscale{0.8}
\plotone{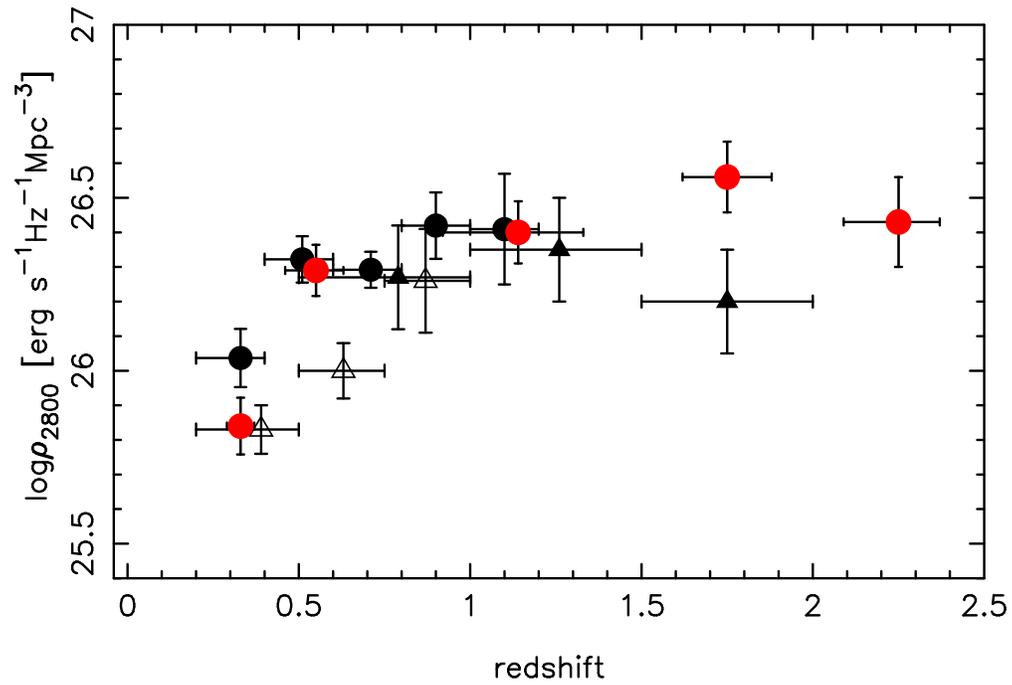}
\caption{Evolution of the rest-frame 2800\AA~luminosity density
in the GOODS CDF-S is shown with red dots. Luminosity densities
are not corrected for dust extinction. For comparison, we also plot the
2800\AA~luminosity density from  COMBO-17 (Wolf et al. 2003; black dots), 
Lilly et al. (1996; open triangles) and Connolly et al. (1997; filled triangles).
Results from literature have been converted to the adopted cosmology.
\label{fig3}}
\end{figure}

To derive the UV LD over all magnitudes, we approximate 
the LF in each bin by the derived Schechter function 
parameters. The LD is given by
\begin{equation}
\rho_{\nu} = \int L_{\nu}\phi(L_{\nu},z)dL_{\nu}=\Gamma(2+\alpha)\phi^*L_*.
\end{equation}
The resulting LDs are listed in Table 1. In 
Figure \ref{fig3}, we plot the 2800\AA~LDs
as red dots. We note an increase in the LD
by a factor $\sim 4$~between redshift $z\sim 0.3$~and $z\sim 1$. At higher
redshift, the evolution is consistent with being flat. For comparison, 
we also plot the LDs from the
Canada France Redshift Survey (CFRS; Lilly et al. 1996), 
HDF-N (Connolly et al. 1997) and COMBO-17 (Wolf et al. 2003). There is in 
general an excellent agreement at overlapping redshifts between the GOODS and 
the COMBO-17 results, however, at $z\sim 0.3$, we find a somewhat lower value, 
suggesting a stronger evolution to $z\sim$1. At $z\gsim$1.5, we find a 
LD that is a factor $\sim$2 higher than the results given in 
Connolly et al. (1997). We note, however, that part of the difference
between results may be due to clustering variance, especially affecting surveys
with small areas such as the HDF-N (Connolly et al.) which only 
covers $\sim$5 sq. arcmin. The deviating point from Lilly et al. at $z\sim$0.6
could also be due to a combination of statistical scatter and clustering 
variance. 

\begin{figure}
\epsscale{0.8}
\plotone{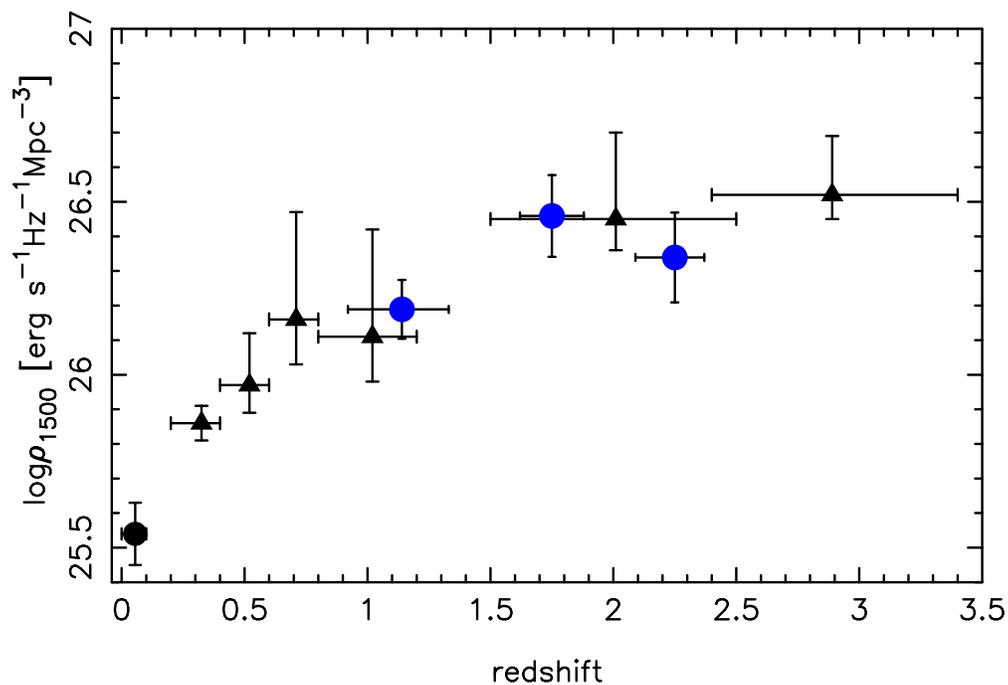}
\caption{Evolution of the rest-frame 1500\AA~luminosity density
in the GOODS CDF-S is shown with blue dots. Luminosity densities
are not corrected for dust extinction. For comparison, we also plot the
1500\AA~luminosity density from GALEX (Schiminovich et al. 2005, 
triangles and Wyder et al. 2005, black dot).
Results from literature have been converted to the adopted cosmology.
\label{fig4}}
\end{figure}

The rest-frame 1500\AA~LD is plotted in Figure \ref{fig4}~as
blue dots. There is a trend showing an increase in the LD between
redshifts $z\sim 1$~and $z\sim 2$. However, our highest redshift point
again suggests a flattening in the evolution. In the Figure, we also
plot recent measurements from GALEX (Schiminovich et al. 2005; triangles
and Wyder et al. 2005; black dot). Together, these measurements
depict a scenario where the LD evolves rapidly between $z=0$~and
$z\sim 1$, thereafter the evolution is somewhat less steep and 
at $z\sim 2-3$, the evolution flattens out, or may even turn over. 

\section{Deriving the SFR from UV luminosity densities}
Since the UV luminosity is mainly produced by short-lived O and B stars,
it is closely related to the ongoing star formation. Using results from
stellar synthesis codes and assumptions on the past star formation history, 
it is possible to derive a conversion factor
relating the UV luminosity and the ongoing SFR.
For example, Madau et al. (1998) present relations between the 1500\AA~
and 2800\AA~luminosities and the SFR assuming an exponentially declining 
star formation history, SFR$\propto$exp(t/$\tau$). They find that 
the relation is quite insensitive to the past star formation
history when varying $\tau$~in the interval 1 to 20 Gyr. 
We have examined the results in Madau et al. and confirm
that at 1500\AA, the relation holds and is independent from
the star formation history while at 2800\AA, the conversion
factor varies by $\sim$30\% depending on $\tau$. The larger spread
at 2800\AA~is due to the fact that more long-lived A and F stars
are significant contributors here in addition to O and B stars.

To investigate the relation between luminosity density, ongoing
star formation and star formation history averaged over an ensemble
of galaxies, we use 
two different models. In the first, we assume
a constant SFR($z$). This should be approximately true at $1.5 < z < 6$~where
we know from independent measurements (e.g., Giavalisco et al. 2004a) that
the star formation rate is consistent with being flat. As our second
model, we use the fit to a number of SFR measurements presented in 
Giavalisco et al. and parameterized in Strolger et al. (2004),
\begin{equation}
SFR(t)=a(t^be^{-t/c}+de^{d(t-t_0)/c}),
\end{equation}
where $t$ is the age of the universe (in Gyr), $t_0$=13.47 Gyr, $a$=0.021, 
$b$=2.12, $c$=1.69, and $d$=0.207. In both models, we assume an onset of the 
star formation at $z=6$, the results are, however, not sensitive to the 
exact choice of this redshift. Also, the conversion factor is only dependent 
on the shape of the SFR history and not the absolute normalization (i.e., 
parameter $a$~is arbitrary in the equation above). This also means that 
results are independent of any constant amount of dust extinction.
Input are also
GALEXEV stellar population synthesis models (Bruzual \& Charlot 2003), giving 
the age luminosity evolution for a simple stellar population (SSP) at 
different UV luminosities. For these models, we assume solar 
metallicity and two different IMFs, a standard Salpeter IMF (Salpeter 1955)
and a Chabrier IMF (Chabrier 2003). The relation between ongoing SFR and
UV luminosity is finally derived by convolving the SSP
results with the SFR history
\begin{equation}
k_{\nu}(t)=\int^t_{t_{z6}}SFR(t^{\prime})\times l_{\nu}(t-t^{\prime})dt^{\prime}/SFR(t),
\end{equation}
where $l_{\nu}(t^{\prime})$~is the flux at time $t^{\prime}$~after an 
instantaneous burst of the SSP and $t_{z6}$~is the age of the universe 
at $z=6$. The conversion factor can thereafter be used to calculate the 
ongoing SFR from the observed UV luminosity density using
\begin{equation}
L_{\nu}=k_{\nu} \frac{{\rm SFR}}{M_{\odot}~{\rm yr}^{-1}}{\rm ergs}~{\rm s}^{-1}~{\rm Hz}^{-1}.
\end{equation}

\begin{figure}
\epsscale{0.8}
\plotone{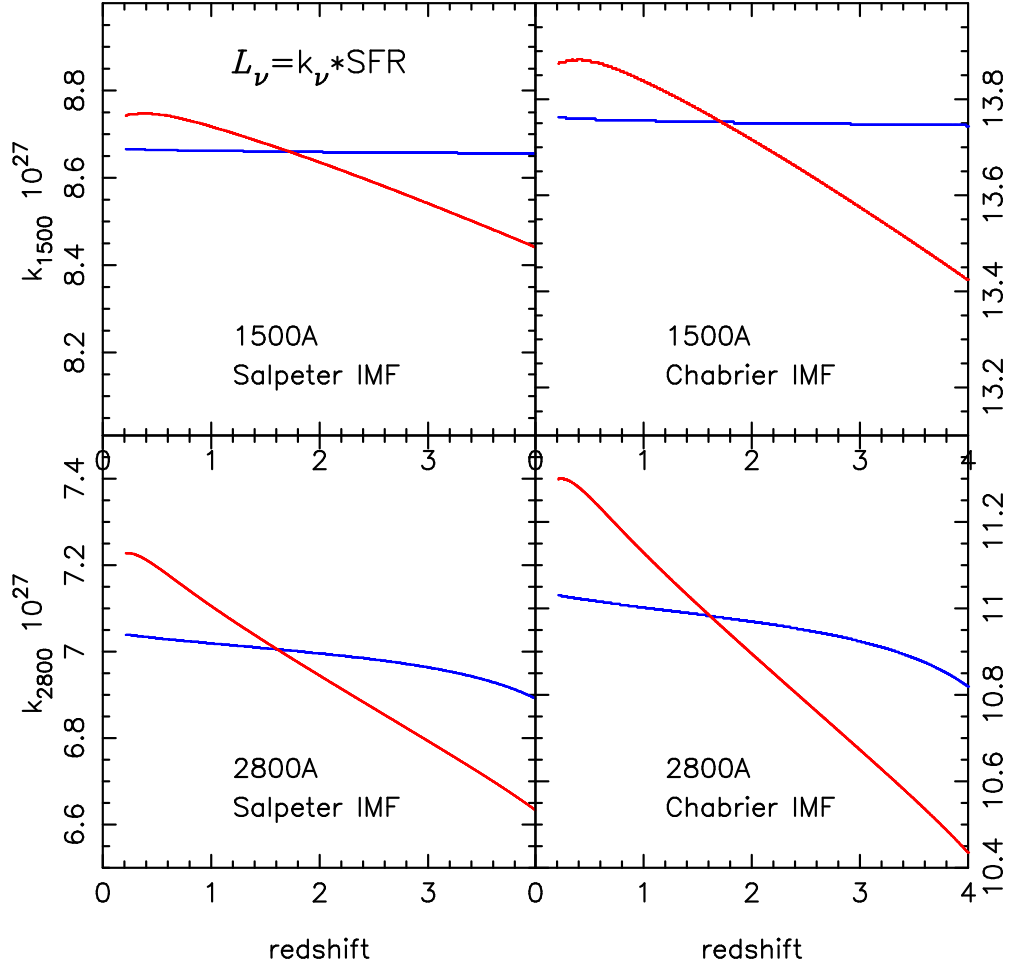}
\caption{Relation between UV luminosity density and ongoing SFR for two
different assumptions on the past star formation history. Our first model
assumes a constant SFR history (blue lines), while the second assumes
an evolving SFR(z) and is taken from Strolger et al. (2004). Top panels
show results at 1500\AA, while bottom show results at 2800\AA. We use
solar metallicities and two different IMFs (Salpeter in left panels and
Chabrier in right panels). Conversion factors are given in units 10$^{27}$.
\label{fig5}}
\end{figure}

In Figure \ref{fig5}, we show the resulting conversion factors between
UV luminosity and ongoing SFR at rest-frame wavelengths 1500\AA~and 2800\AA.
Note that we at this stage do not include any corrections for
dust extinction.
For the constant SFR scenario (blue lines in Figure \ref{fig5}), we find 
that the conversion factor is close
to independent of redshift. This is expected since we in this case expect
that the fraction of the UV luminosity coming from ongoing star formation
remains constant. At 2800\AA, however, there is a slight increase in 
$k_{2800}$~with time due to the build-up of old stars with residual 
2800\AA~luminosity. 
For the more realistic SFR history (red lines in Figure \ref{fig5}), 
we note a clear, albeit small, redshift dependence
on the relation between UV luminosity and ongoing SFR. 

At 1500\AA, 
the conversion factor increases by $\sim 2.3$\% between $z=3$~and $z=0.3$~
(similar for both IMFs). The evolution is, as expected, stronger at 2800\AA, 
where we note an increase by $\sim$6\% over the the same redshift range.
As a consistency check, we calculate what fraction of the observed 2800\AA~
luminosity density comes from galaxies that are best-fitted by an early-type
SED, i.e., representing non-star forming galaxies. We do this by fitting 
a Schechter function to early-type population separately in the two 
lowest redshift bins. We find that $\sim 10\pm 4$\%~of the luminosity density
comes from this population, consistent with the theoretical predictions.
At higher redshift, the statistics are too poor to allow a fit to the
early-type population separately.

\begin{figure}
\epsscale{0.8}
\plotone{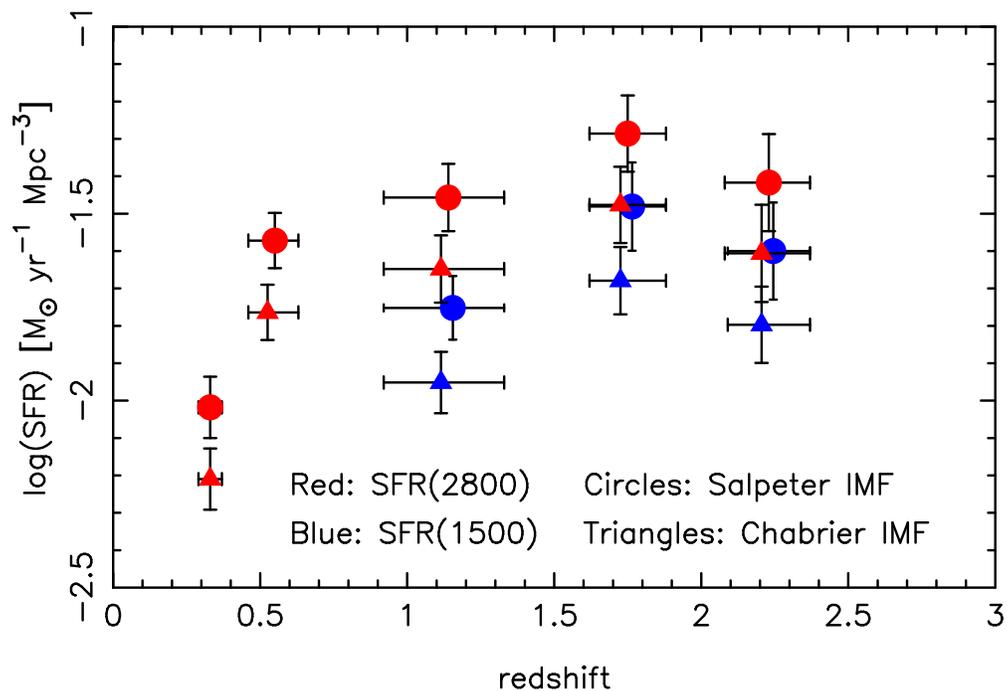}
\caption{SFRs derived from 1500\AA~and 2800\AA~luminosity densities
(blue and red dots, respectively). We show results for both a Salpeter
IMF (circles) and a Chabrier IMF (triangles). The SFR derived from the
2800\AA~luminosity is a factor $\sim$1.7 higher compared to the 1500\AA~
derived SFR for both choices of IMF. 
\label{fig6}}
\end{figure}

We use the conversion factors derived from the realistic SFR history
to derive the ongoing SFR for the GOODS dataset. Results are shown 
in Figure \ref{fig6}. 
For both IMFs, we find that the SFR derived from the 2800\AA~luminosity
is a factor $\sim 1.7$~higher compared to the SFR derived from the 1500\AA~
luminosity. Since we have accounted for the fraction of the 2800\AA~
luminosity that comes from a 'non-star forming' population, the remaining
discrepancy should have other causes. 
First, an IMF with a different slope at high masses compared to
the standard Salpeter IMF and Chabrier IMF, could change the 
relation between UV luminosities and SFR, and therefore explain some
of the difference.
Second, 
it is expected that a large fraction of the UV luminosity is obscured by
dust extinction. Since the extinction is more severe at shorter 
wave-lengths, this could explain the difference in the SFRs 
derived from 1500\AA~and 2800\AA~luminosities.

Assuming a Calzetti attenuation law (Calzetti et al. 2000) and a Salpeter IMF,
we fit the amount of dust needed to bring the SFRs from 1500\AA~and 2800\AA~to
agreement and find that a mean E(B--V)=0.20$\pm0.05$. This is equivalent to an 
extinction corrected SFR
a factor $\sim 6.5$~higher at 1500\AA~and a factor $\sim 3.7$~higher 
at 2800\AA, compared to the uncorrected SFRs.  

The extinction correction we find here is higher compared
to the median E(B--V)=0.15 found by Shapley et al. (2001), who 
derive the extinction from optical and near IR photometry of a set 
of $z\sim 3$~Lyman break galaxies (also assuming a Calzetti attenuation 
law and a Salpeter IMF). However, the Shapley et al. (2001) results are based
on Bruzual \& Charlot (1996) models. Using the more recent models in Bruzual
 \& Charlot (2003) results in a median E(B--V)=0.20 for the same galaxy
sample (Shapley, private communication), which is consistent with what found 
here.

The correction at 1500\AA~is
somewhat larger than the factor $\sim 5$~derived from the GALEX data
by Schiminovich et al. (2005). Our results suggest that 
the extinction corrected SFR is up to $\sim 50\%$~higher than 
previously derived extinction corrected rates based on UV luminosities, 
e.g., Giavalisco et al. (2004a). 

\begin{figure}
\epsscale{0.8}
\plotone{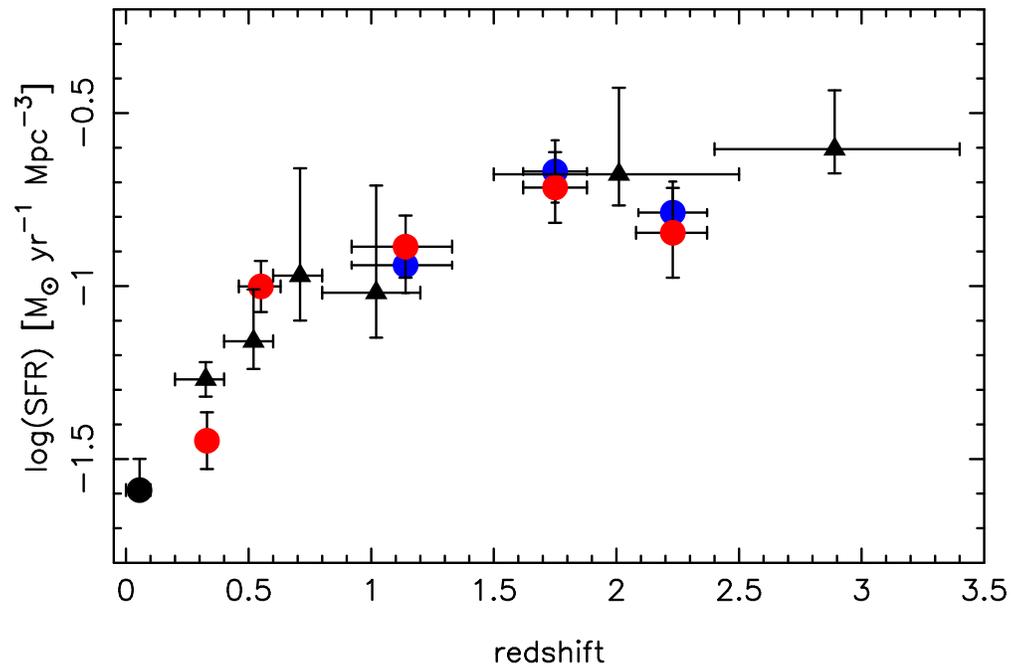}
\caption{Extinction corrected SFRs derived from 1500\AA~and 2800\AA~luminosity densities
(blue and red dots, respectively), assuming a Salpeter IMF. Also shown are SFRs
calculated from 1500\AA~LD in Wyder et al. (2005; black dot) and Schiminovich
et al. (2005; triangles).
\label{fig7}}
\end{figure}

In Figure \ref{fig7}, we show the extinction corrected SFRs 
derived from both the 1500\AA~and 2800\AA~luminosity densities.
For the two lowest redshift bins, we derive the SFR from the 
2800\AA~luminosity density assuming the same extinction correction 
as derived from the results at $z>1$.
Our results support a steep increase in the SFR between
$z=0$~and $z\sim 1$. Compared to the local measurement by Wyder et al. (2005),
we find an increase in the SFR by at least a factor $\sim 5$ between $z=0$ 
and $z\sim 1.5$. 
At redshifts $z\gsim 1.5$, there is a mild increase in the SFR, however, we can
not exclude a flat rate over the range $1\lsim z\lsim3$.
 
In the Figure, we also plot SFRs based on 1500\AA~luminosity densities 
from Schiminovich et al. (2005; triangles). We have used the corrections 
derived
here to calculate the SFRs from the LDs. We note that the agreement 
is good, and only for the lowest redshift point in the GOODS sample 
is the deviation outside the one sigma error bars. 
\section{Morphology of star forming population}
\subsection{Determining galaxy morphologies}
We use GALFIT (Peng et al. 2002) to determine the morphology at 
rest-frame 2800\AA~for 
the objects in the $HST$~ACS $BViz$~bands. 
We use unconvolved images when deriving morphological catalogs.
These are thereafter matched to the photometric catalogs
(based on convolved images).
We do not include the ground-based $U$-band data in the morphological
investigation (i.e., the $z\sim 0.33$~redshift bin)
due to the lower resolution and broader psf in this band. 
The light distribution of each object is fit with a S\'{e}rsic (1968) 
radial profile
\begin{equation}
\Sigma(r)=\Sigma_e e^{-\kappa[(r/r_e)^{(1/n)}-1]},
\end{equation}
where $n$~is the S\'{e}rsic index, $\kappa$~is defined as $\kappa$=(2$n-$0.331), $r_e$~
is the effective radius containing half of the total galaxy light and
$\Sigma_e$~is the surface brightness at $r_e$. In the fitting we use
a psf derived from stars in the field.
A pure exponential profile has $n=1$, while a 
de Vaucouleurs profile has $n=4$. We adopt a division between ``disk''-objects 
with $n<2.5$, and 
``bulge''-objects with $n>2.5$, which is the same criterion as used by e.g.,
Barden et al. (2005) and Ravindranath et al. (2006). 
Bulge objects should
mostly consist of ellipticals, while disk objects are foremost spirals and 
irregulars. We note, however, that at high redshifts and short rest-frame 
wavelengths, we do not expect the galaxy population to follow the Hubble 
sequence as well as the case is locally. 

In order to derive S\'{e}rsic index and galaxy radius it is necessary that
a major fraction of the galaxy luminosity comes from a main source for which
a center can be well defined. When going to higher redshifts, and in particular
in rest-frame UV, it is expected that galaxy morphology becomes more
disturbed with the possibility of multiple peaks in the light distribution. 
This could affect the measured properties, making it difficult to interpret 
e.g., S\'{e}rsic index and galaxy size. In order to investigate the galaxies 
included in this investigation, we visually inspected representative 
samples of galaxies in all redshift bins, including both ``disk''- 
and ``bulge''-objects. We find that ``bulge''-objects in all bins are well 
represented by symmetric and single peaked objects. At all redshifts, 
``disk''-objects are, as expected, more irregular in shape. However, we find 
that the sample we select is clearly dominated by galaxies with a main 
luminous component for which a center can be defined. Therefore, we trust 
that the S\'{e}rsic index and galaxy radius derived reflects the 
morphological properties of the galaxies.

The reason for the relatively well behaved morphologies of these galaxies
can be attributed to the selection, where we only include intrinsically
bright objects. For these, we do not expect that star forming lumps or
trails of tidal interactions affect the overall morphology more than 
marginally in most cases. For fainter objects, the situation should be 
different where secondary star forming regions can cause multiple peaks 
with similar luminosity. Although we can calculate S\'{e}rsic index 
also for these objects, we do not include them
in our investigation since the interpretation of these is more uncertain.

\begin{figure}
\epsscale{0.7}
\plotone{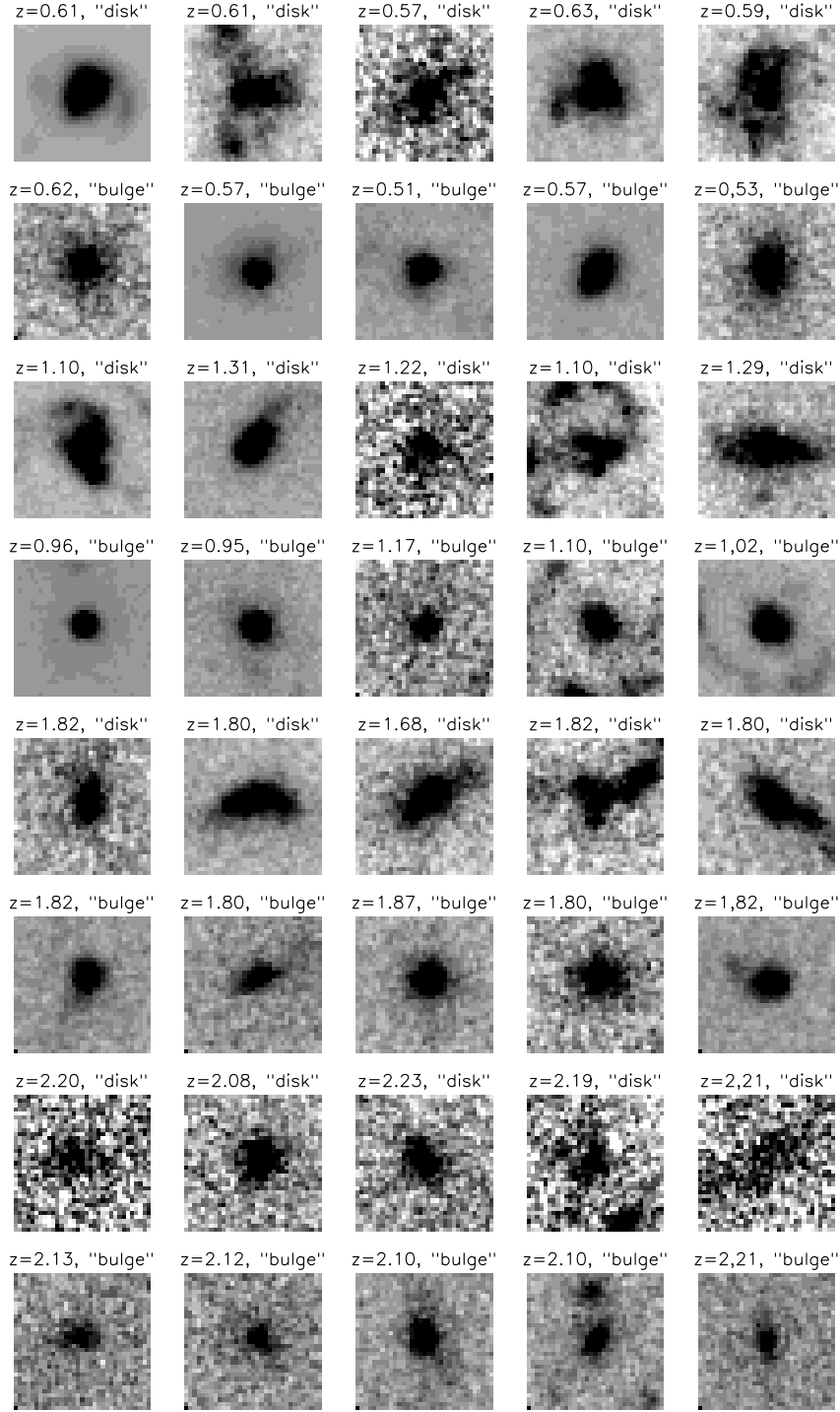}
\caption{A representitative sample of five ``Disk''- and five ``Bulge''-objects
in each of the four redshift bins at rest-frame 2800\AA. ``Disk''-objects are 
defined as objects with Sersic index $n<$~2.5, while ``Bulge''-objects have
$n>$~2.5. The size of each 
postage stamp is 1$\times$1 arcsec (with a pixel scale 0.03 arcsec/pixel), where
1 arcsec corresponds to 6.4, 8.2, 8.5, and 8.2 kpc at redshifts $z\sim 0.55$, 
$z\sim 1.14$, $z\sim 1.75$, and $z\sim 2.23$, respectively. The stretch of 
the gray-scale in each postage stamp depends on the 
brightness of the objects, therefore images of fainter galaxies have
a higher level of graininess.
\label{fig8}}
\end{figure}

In Figure \ref{fig8}, we show representitative samples of ``disk''- and
``bulge''-objects in each redshift bin. We have chosen these randomly 
among galaxies brighter than $M_C$ (described below). For the ``bulge''-objects,
the selected galaxies represent 100\%, 12\%, 9\%, and 24\% of the full
sample in the four redshift bins, respectively (from low to high redshift).
For ``disk''-objects, the corresponding fractions are 10\%, 3\%, 2\%, and 17\%,
respectively. As discussed above, the figure shows that the ``bulge''-objects
are more symmetric and mostly consist of a single central object while
``disk''-objects are more irregular in shape. 

To compare the same population in each band, we use an
absolute magnitude cutoff that is related to the characteristic magnitude 
in each band; $M_C=M^*+\Delta M$ where $\Delta M$~is derived from the 
completeness limit in the highest redshift bin, $M_C=$--19.75.  
We find $\Delta M=0.5$, i.e., in each band we select galaxies with
$M<M_C=M^*+0.5$. The completeness
in determining the S\'{e}rsic index for galaxies brighter than the
magnitude limit is 80\%, 94\%, 94\%~and 93\% in the $B$, $V$,
$i$~and $z$-bands, respectively. There are two reasons for the
incompleteness. First, since S\'{e}rsic indexes are derived using the 
unconvolved images, while photometry is derived using convolved,
there are cases where objects do not match between catalogs.
This is because multiple nearby objects in the unconvolved image may be
merged into a single object in the convolved image. For these objects we do
not assign a S\'{e}rsic index. About 75\% of the objects without
determined S\'{e}rsic index belong to this group.
Second, for the remaining 25\%, the S\'{e}rsic fitting algorithm did 
not converge, mainly due to the faintness of the objects.

\subsection{Relation between morphology and spectral type}
Besides deriving the {\it morphological} characteristics of the galaxy 
sample using the
S\'{e}rsic index, we also derive the {\it spectral} type describing the 
overall
shape of the galaxy SED. Spectral types are given by the best fitting
template SED derived from the photometric redshift fitting. We number
spectral types 1-5, with E=1, Sbc=2, Scd=3, Im=4 and starburst=5 
(for both starburst templates used). We also include intermediate types 
which are interpolations between subsequent templates. 
We divide the galaxies into two spectral types: early-type galaxies
(dominated by E spectrum, type $<$ 1.5) and late-type galaxies
 (type $>$ 1.5).

Using our full data set, we calculate the median S\'{e}rsic index for the 
early-type and late-type populations. We find that the early-type population
has a high median S\'{e}rsic index, $n=5.1$, indicating that galaxies with 
early-type spectra are bulge dominated. The median S\'{e}rsic index for the 
late-type population is $n=1.0$, consistent with a disk-dominated (spiral) 
galaxy.

If we assume that the S\'{e}rsic index in general follows the spectral type 
of the galaxy, we can use this assumption to correct for the incompleteness
in derived S\'{e}rsic indexes described in \S 5.1. We do this by placing
galaxies with early-type spectra in the high S\'{e}rsic sample ($n > 2.5$),
while later type galaxies are placed in the low S\'{e}rsic sample ($n < 2.5$).
Note, however, that there is a large dispersion in the correlation between 
galaxy spectral and morphological type, and we find that $\sim 23\%$~of the 
late-type galaxies have high S\'{e}rsic indexes ($n>2.5$), suggesting that 
they are bulge dominated, and that 
a similar fraction of the early-type galaxies have low S\'{e}rsic index. 

\subsection{Distribution of morphological types}
\begin{figure}
\epsscale{0.8}
\plotone{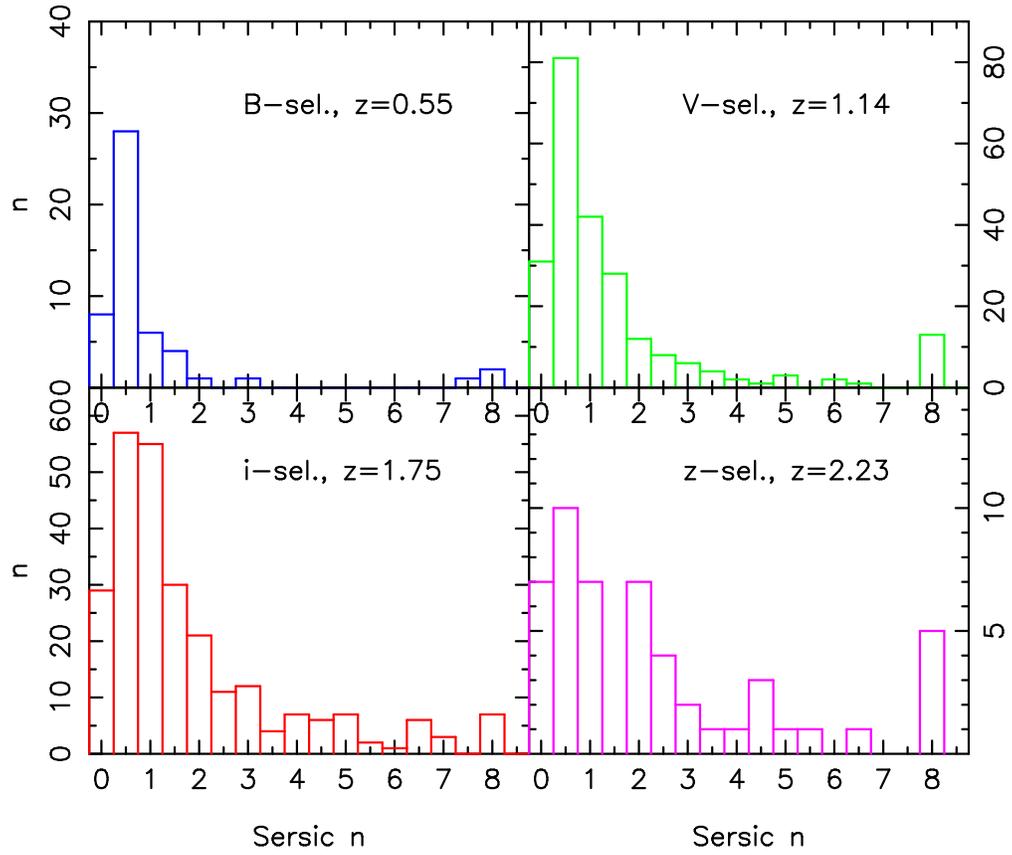}
\caption{Distribution of S\'{e}rsic indexes in the four redshift 
bins. Plotted are galaxies to an absolute magnitude limit $M_C$, as 
described in the text.
\label{fig9}}
\end{figure}

In Figure \ref{fig9}, we show the distribution of S\'{e}rsic indexes in each
redshift bin for galaxies brighter than the limit $M_C$.
The figure shows that the rest-frame 2800\AA~morphology is dominated by
low S\'{e}rsic indexes, i.e., ``shallow'' disk dominated or irregular systems. 
This is natural since star 
formation is expected to occur in these systems. If we calculate the 
fraction of galaxies that are bulge dominated ($n>2.5$), we find that
8\%, 15\%, 24\% and 31\% of the galaxies belong to this group at $z\sim 0.55$,
$z\sim 1.14$, $z\sim 1.75$, $z\sim 2.23$, respectively. 
This suggests that the fraction of
bulge dominated systems (in rest-frame 2800\AA) increases with redshift. 

To investigate the effect on results due to incompleteness in determined 
S\'{e}rsic indexes, we assign a high S\'{e}rsic index ($n>2.5$) to early-type
galaxies and a low  S\'{e}rsic index ($n<2.5$) to late-type galaxies, as
discussed in \S 5.2.  
After making the correction, we find that 9\%, 14\%, 22\%, and 29\% of the
galaxies have high S\'{e}rsic indexes, in the four bins, respectively. The 
trend in the evolution remains, suggesting that the incompleteness only 
marginally affects results.

We have also examined if there is a bias in the derived S\'{e}rsic 
indexes as the mean apparent magnitude of the galaxies gets fainter at 
higher redshift. We investigate this by adding 
simulated galaxies
with known radial profiles (exponential and de Vaucouleurs) spanning a
range of magnitudes and radii to the real ACS $z$-band image using the IRAF 
task MKOBJECTS. We thereafter run GALFIT and determine the fraction of the 
galaxies for which we correctly recover the input profile. At magnitudes 
$z<24$, we find that $>99$\%~of both galaxy types have correctly determined 
profiles. At the faintest magnitudes included in this investigation 
$z\sim 25$, we find that 95\%~of the input exponential profiles are recovered 
as disk galaxies, while 88\%~of the de Vaucouleurs profiles are correctly 
assigned a bulge profile. We therefore conclude that the magnitude bias when 
determining S\'{e}rsic index should not be severe. 

Next, we investigate how the relation between morphological and 
spectral types evolves with redshift. Deriving the median S\'{e}rsic index
in the four redshift bins we find $<n>$=0.49, 0.76, 1.27, and 1.34 at
$z\sim$0.55, $z\sim$1.14, $z\sim$1.75, and $z\sim$2.23, respectively. 
This is consistent with 
the general trend found above showing that the fraction of high S\'{e}rsic 
index objects increases with redshift. Calculating the median spectral
type in the four bins, we find $<$type$>$=3.3, 3.7, 5.0, and 5.0, 
respectively. Overall, this shows that the galaxy population producing the 
2800\AA~luminosity mainly consists of galaxies with
late-type SEDs which get bluer at higher redshift. This is
contrary to the overall correlation between types where later spectral types
in general have lower S\'{e}rsic indexes. Our result that the S\'{e}rsic index
increases with redshift, while the galaxies at the same time get bluer,
again suggests that a larger fraction of the star formation at high 
redshift occurs in more concentrated bulge-dominated objects compared 
to low redshifts.          

\begin{figure}
\epsscale{0.8}
\plotone{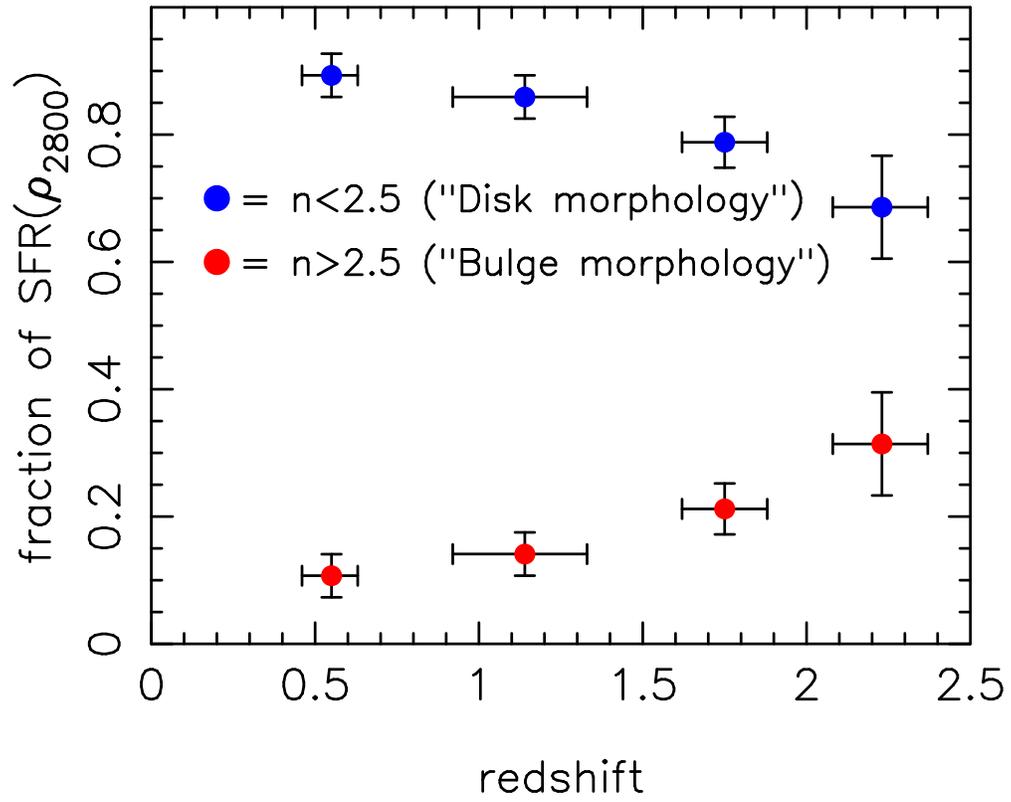}
\caption{The fraction of the star formation rate originating in
bulge-dominated ($n>2.5$) systems (red dots) and disk systems (blue dots),
respectively.
\label{fig10}}
\end{figure}

So far, we have investigated the fraction of the number of galaxies
belonging to the different morphological types. We now estimate the fraction
of the rest-frame 2800\AA~luminosity emitted by disk- and bulge-dominated
systems. We do this by summing the emitted flux in each redshift 
bin and thereafter determining what fraction comes from bulge systems. 
Using the magnitude limit $M_C$, we find fractions 0.11$\pm0.04$, 
0.14$\pm0.04$,
0.21$\pm0.04$~and 0.31$\pm0.08$, 
of the luminosity is emitted in bulge ($n>2.5$) systems at $z\sim 0.55$, 
$z\sim 1.14$, $z\sim 1.75$, and $z\sim 2.23$, respectively. These results 
are plotted in Figure \ref{fig10}. The trend we find is inconsistent with
a flat non-evolution scenario at a $\sim 3\sigma$~level. Note that we
do not extrapolate the LF to faint magnitudes. If the faint-end slopes
of the disk and bulge population differ, or evolve differentially with time, 
then conclusions could be different at the faint end.

Similar results, showing an increased fraction of the SFR occurring
in spheroids/ellipticals to $z\sim 1$, are presented by Menanteau et al. 
(2006), who find that the fraction of the SFR occurring in ellipticals at 
rest-frame $B$~is
$\sim0.1$~at $z\sim0.5$~and $\sim0.16$~at 
$z\sim1.1$. At higher redshift ($z\gsim 1.2$), the fraction decreases
in Menanteau et al., in contrast to the continuous increase suggested in this
investigation.

At lower redshifts, Lauger et al. (2005) find a steep increase
in the number of faint blue bulge-dominated galaxies between $z\sim 0.15$~and
$z\sim 1.1$~in rest-frame $B-$band. The emergence of this population could 
be related to the morphological evolution at higher redshifts seen in 
this investigation.

Conselice et al. (2005) investigate the relation between rest-frame $B$-band
morphology and star formation in the HDF-N. They find that star formation is 
higher in more massive and early-type galaxies at $z>1$~compared to lower
redshift. This is consistent with what is found in our investigation.
Our results are also in agreement with the result of
Menanteau et al. (2001), who use $HST$~data 
with $I_{814W}<24$~mag from the Hubble Deep Fields N and S
and find that a significant
fraction ($\sim 30\%$) of intermediate redshift ($z\sim 1$) 
spheroids/ellipticals have blue colors.
 
Furthermore, Lotz et al. (2006) find a fraction
$\sim$30\% spheroids at $z\sim 4$~and $\sim$15\% spheroids at $z\sim 1.5$. 
This is consistent with the trend found in this investigation showing a 
decrease
in spheroid fraction at lower redshifts. The fractions found are also
reasonably consistent with the numbers we find. Note, however, that
Lotz et al. measure spheroid fraction at a different rest-frame 
(FUV, $\lambda < 2000\AA$) and use a different method for defining spheroids, we could therefore expect some differences.

Most of the stellar mass in ellipticals is expected to have
assembled by $z\sim$1 (Bundy et al. 2005). The increased fraction
SFR in bulge systems at higher redshifts we observe could therefore
be the sign of the build up of ellipticals at $z>1$, and the 
subsequent decrease in SFR reflects a more passive evolution
at $z<1$. This is similar to the scenario Menanteau et al. (2001) 
suggest where the blue galaxies they
observe at $z\sim$1 are old elliptical systems 
undergoing recent star formation induced by mergers or inflow of material. 
However, we can not exclude that some part of the bulge systems we
observe at high redshift may evolve to the bulges of todays population
of old giant spirals.

To summarize, we find an increasing fraction of 
galaxies at higher redshifts to be bulge-dominated in the rest-frame 2800\AA. 
This suggests that a significant amount of star 
formation occurs in these objects at high redshift. At least part of 
the increase in SFR should be due to the build up of todays ellipticals
by infall or mergers at $z\gsim$1. 

\subsection{Size-redshift relation at 2800\AA}
Simple scaling models for the expected redshift evolution of galaxy sizes 
in the hierarchical model is 
presented by e.g., Fall \& Efstathiou (1980), Mo et al. (1998) and 
Bouwens \& Silk (2002). The scale 
length of a spiral galaxy, $R_s$, is assumed
to be proportional to the virial radius, $R_{\rm vir}$~and therefore
related to the virial mass via
\begin{equation}
R_s\propto R_{\rm vir}\propto V_{\rm vir}/H(z) \propto {\rm M_{vir}}^{1/3}/H(z)^{2/3},
\end{equation}
where
\begin{equation}
H(z)=H_0[\Omega_M(1+z)^3+\Omega_k(1+z)^2+\Omega_{\Lambda}]^{1/2}
\end{equation}
and $V_{vir}$~is the circular velocity at $R_{vir}$.
For fixed circular velocity, we therefore expect $R_s\propto H(z)^{-1}$,
while for fixed mass we expect $R_s\propto H(z)^{-2/3}$. In the case of
a sample with fixed rest-frame luminosity, the behavior depends on the
evolution of the mass-to-light ratio and is expected 
to be intermediate between these relations (Bouwens \& Silk 2002; Ferguson
et al. 2004).

\begin{figure}
\epsscale{0.8}
\plotone{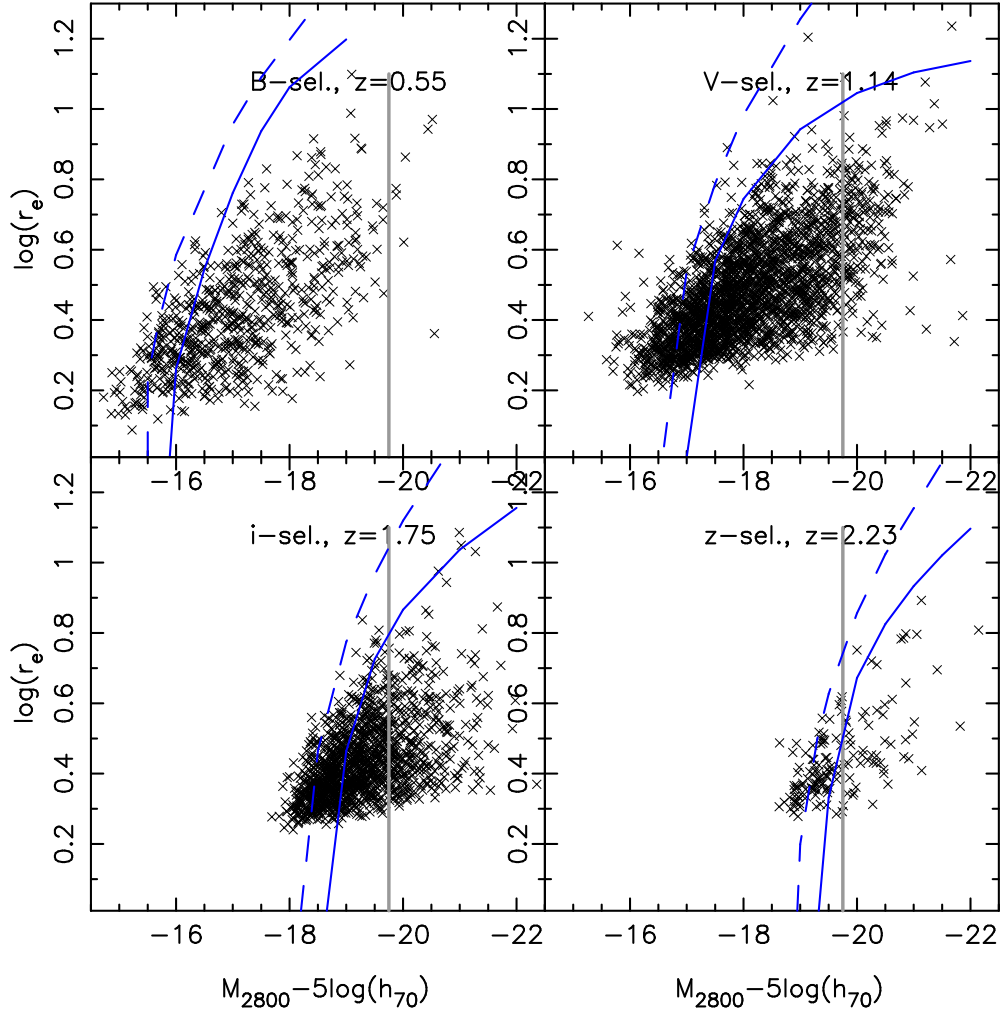}
\caption{Measured half-light radius vs. absolute magnitude in the four 
redshift bins. In each bin we show the 99\% (solid line) and 50\% (dashed line)
completeness limits derived from simulations. Gray line shows the adopted
selection criteria.  
\label{fig11}}
\end{figure}

\begin{figure}
\epsscale{0.8}
\plotone{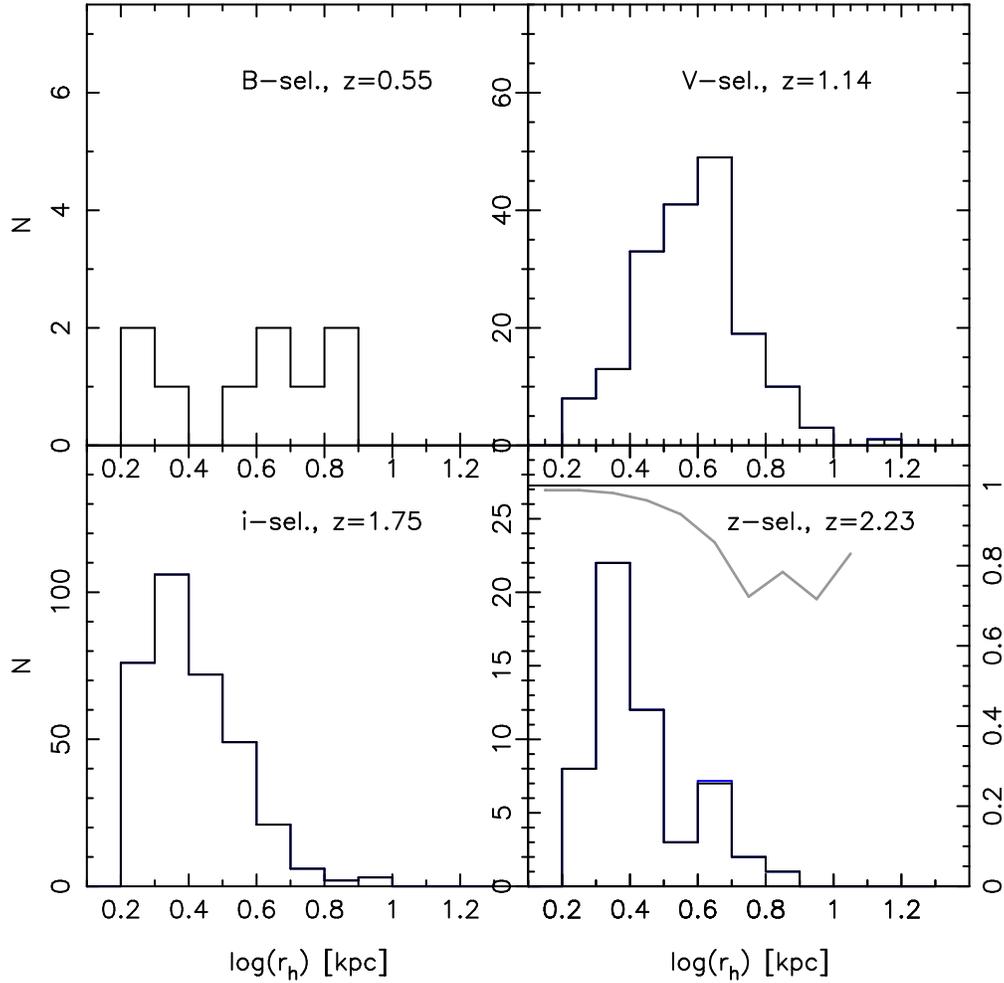}
\caption{Distribution of half-light radii for galaxies with $M_{2800}<-19.75$~
in the four redshift bins. In the highest redshift bin, we also plot the expected
completeness as a function of radius based on simulations where we shifted the galaxy 
population at $z\sim 1.1$~(second bin) to the highest redshift bin, thereafter
we determine the recovered fraction of galaxies as a function of radius.
\label{fig12}}
\end{figure}

When investigating the size evolution, we want to compare galaxies 
within the same range of e.g., absolute magnitudes or masses
(compared to the above investigation where we examine the brightest 
population in each bin).
We investigate the completeness in each bin by adding simulated galaxies 
to the real images. The simulated galaxies span a wide range of absolute
magnitude and size (i.e., surface brightness). In Figure \ref{fig11}, 
we plot the half-light radius ($r_h$, as measured by SExtractor) at 
rest-frame 2800\AA~of the observed galaxies as a function of
absolute magnitude. We only include galaxies with disk morphology
($n<2.5$) since equation (4) implies the formation of a spiral galaxy.
Over-plotted are the 50\% (dashed lines) and 99\% 
(solid lines) completeness limits derived from the simulations. The 
gray line shows the adopted magnitude limit, $M<-19.75$. In 
Figure \ref{fig12}, we plot the distribution of radii for galaxies
brighter than this limit. 

There are only a few objects brighter than our magnitude limit that
fall within the area where 
we expect a completeness between 50\% and 99\% (Figure \ref{fig11}). 
This indicates that we
are not severely affected by incompleteness and should not be 
missing more than a few objects at the most. As a further test,
we have taken the galaxies detected $V$-band ($z\sim 1.1$~redshift bin)
and redshifted their properties (luminosity, size), including effects
of surface brightness dimming.
We thereafter randomly distribute this population on the $i-$~and
$z-$band images and run SExtractor to derive the fraction of 
recovered galaxies.
In the third redshift bin, $z\sim 1.7$, we recover 99\% of the 
redshifted galaxies, with no dependence on the size of the galaxies.
In the highest redshift bin, $z\sim 2.2$, the overall recovered fraction
is 89\%, with with a size dependence suggesting that we detect
$\sim$75\% of galaxies with log(r$_{\rm h}$)$>0.8$~
and $\sim$92\% of galaxies with log(r$_{\rm h}$)$<0.8$. These results
also suggest that we are not severely affected by incompleteness in our
high redshift galaxy samples. The completeness in the highest redshift
bin, derived from these simulations, is shown with the gray line in 
Figure \ref{fig12}. Note that we do not take into account any 
size-redshift relation (i.e., eq. [4]), when redshifting our 
objects from low to high redshift bins. If this relation exists, 
we will be even more complete.  

We use the results from our simulations to account for the incompleteness
by weighting objects within the area between to 50\%~and 99\%~lines 
by the inverse
of the completeness at the particular point in the absolute magnitude-radius
diagram. 
We further note that only a few galaxies survive the selection in 
the lowest redshift bin. This is due to the general fading of the galaxy 
population at lower redshift and the small volume of the lowest bin. 
We keep this bin in our investigation, but note that due to the 
low statistics (large statistical errors), our results regarding
redshift evolution of the galaxy properties do not change whether
we include this bin or not.

As a consistency check on the selected galaxy samples, we calculate 
the median absolute magnitude and 
find $M_{2800}=-20.0$~in the three highest redshift bins, while
the low redshift bin has a median magnitude $M_{2800}=-20.4$.
This reassures us that we are comparing samples with similar magnitudes.

\subsubsection{Size-redshift relation at fixed luminosity}
\begin{figure}
\epsscale{0.8}
\plotone{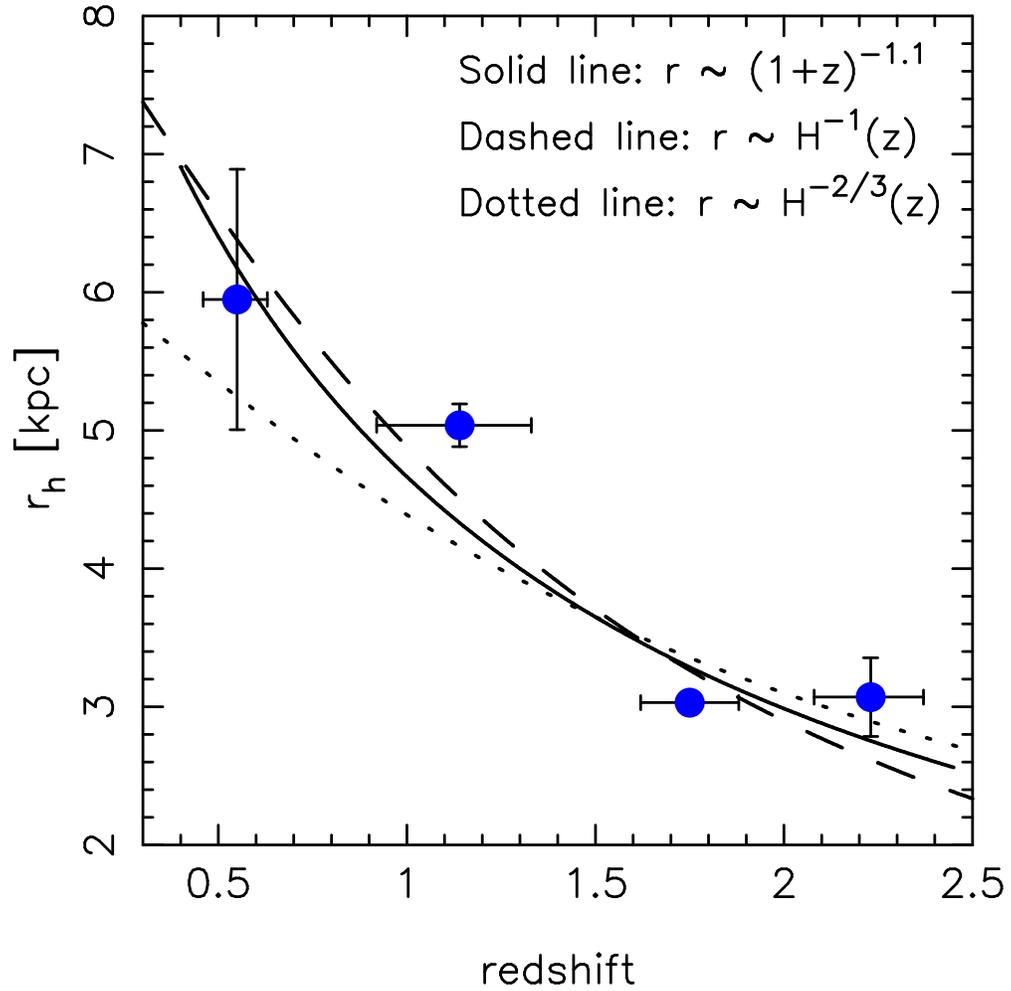}
\caption{Size-redshift relation for disk galaxies selected by absolute magnitude. Blue dots show
median value in each redshift bin. Solid line shows the best-fitting size evolution (1+$z$)$^{-m}$,
where $m=1.1$. Also shown are theoretical curves if sizes evolve as 
$r\propto H(z)^{-1}$~(dashed line) and $r\propto H(z)^{-2/3}$~(dotted line).
\label{fig13}}
\end{figure}

Figure \ref{fig13} shows the size redshift relation for our magnitude
selected galaxy sample. In each bin we plot the median radius in kpc
for our adopted cosmology. Over-plotted are the 
theoretical curves representing the cases where galaxy sizes evolve as  
$R_s\propto H(z)^{-1}$~and $R_s\propto H(z)^{-2/3}$. 
Following the approach in e.g., Bouwens et al. (2004b), we fit our data to the
functional form $r_h \propto (1+z)^{-m}$, and find $m=1.10\pm{0.07}$~
between $1.1\lsim z \lsim 2.2$ (including the $z\sim 0.5$~point do not 
change the derived value of $m$).

If we use the $R_s\propto H(z)^{-1}$~and $R_s\propto H(z)^{-2/3}$~
parameterizations and fit the curves to functional form 
$r_h \propto (1+z)^{-m}$~over the redshift range $z=1.14$~to $z=2.23$,
we get $m=1.3$~and $m=0.9$, respectively. This 
indicates that the evolution we find is somewhat steeper than expected
for a mass selected sample. A possible explanation is that the stellar mass-to-light
ratio decreases at higher redshift. Such scenario is expected as the SFR
in individual galaxies increases at higher redshift. A manifestation of
this is the observed brightening of $M^*_{2800}$.

Compared to previous investigations, our results 
agree well with the evolution $m=1.1\pm{0.3}$~found by Bouwens
 et al. (2005) between redshifts $z\sim6$~and $z\sim2.5$~for a large sample 
dropout galaxies. Even though Bouwens et al. measure the evolution
at a shorter rest-frame wavelength ($\lambda \sim 1350$\AA) and include
all galaxy morphologies, these results 
suggest that the size-redshift correlation continues to $z\sim1$.
Ferguson et al. (2004) investigate the evolution at rest-frame 
$\sim$1500\AA~and find an evolution, $r_h\propto H(z)^{-1}$, 
over the redshift range $1\lsim z \lsim 5$, which is equivalent 
with $m\sim1.4$. This is somewhat steeper than found here, with both
investigations consistent with the $r_h\propto H(z)^{-1}$~relation.

There are also investigations of the size redshift relation 
at optical wavelengths. In rest-frame $B-$band, Papovich et al. (2005) 
find $m=1.2\pm 0.1$~between $z\sim 2.3$~and $z\sim 1$. 
While in rest-frame V-band, Trujillo et al. (2005) find an 
evolution $m=0.9\pm0.2$~
at $1<z<3$~and $m=0.65\pm0.05$~over a larger redshift interval $0<z<3$~for
galaxies with disk morphology ($n<2$).

In combination, these results suggest a similar size-redshift evolution 
over the redshift range $0<z<6$~which is consistent with $m\sim 1$~and, 
as noted by Trujillo et al. (2005), that the
relation is similar at visual and UV rest-frame wavelengths. 

\subsubsection{Size-redshift relation at fixed mass}
Next we investigate the evolution of the size-redshift evolution at
a fixed mass scale. We use the relation between mass and rest-frame 
$V$~magnitude and $B-V$~color given in Bell et al. 
(2003) to determine masses. Rest-frame magnitudes and colors are calculated
using the observed bands that are closest to the redshifted rest-frame $B$~and
$V$-bands, respectively, in combination with K-corrections calculated from 
the best-fitting template SED for each individual galaxy. More details on 
this procedure are given in Dahlen et al. (2005). Derived masses are 
calculated for a Kroupa (2001) IMF (a Salpeter (1955) IMF gives masses 
$\sim$two times higher (Kauffman et al. 2003)). 

We stress that this is not directly a mass selected sample since it  
is based on photometry. Derived masses are subjected
to uncertainties and scatter in these relations. This may be especially severe 
for this investigation since we are using the very bluest galaxy 
population where the relation between mass in luminosity
has a large dispersion.

\begin{figure}
\epsscale{0.8}
\plotone{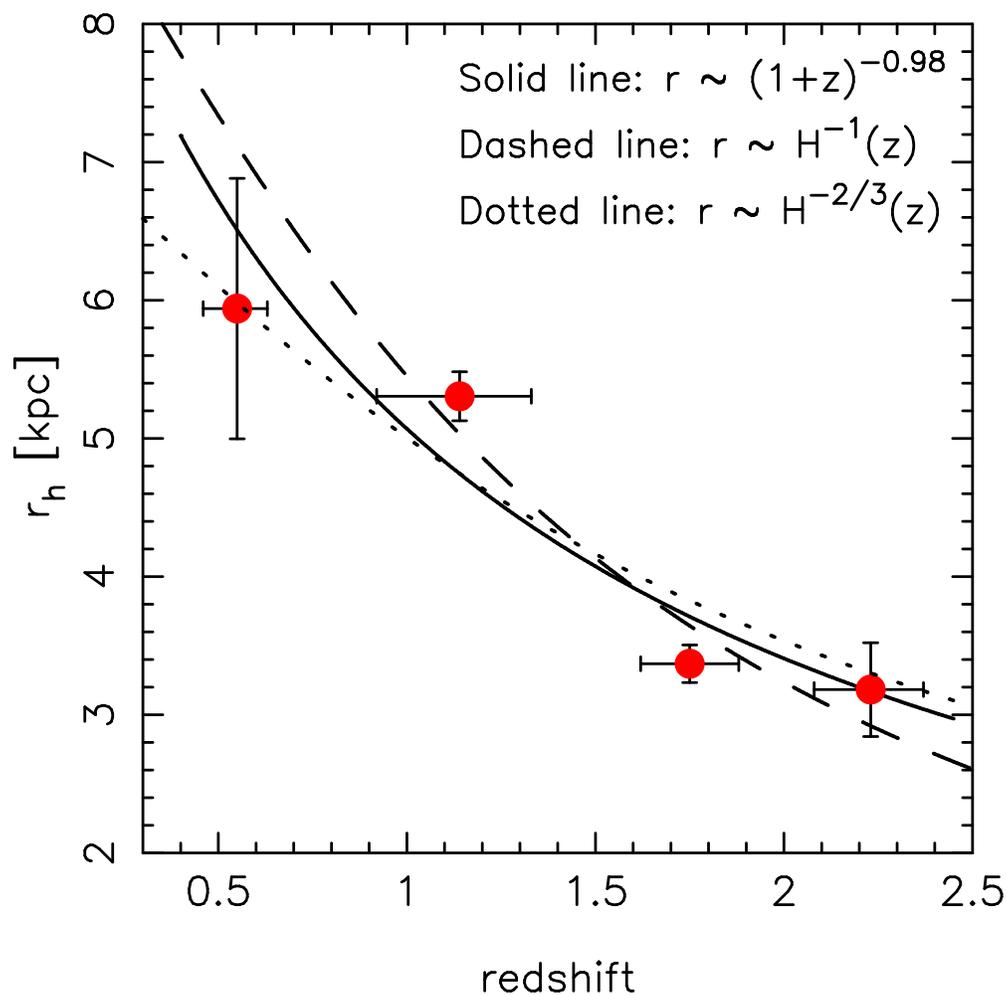}
\caption{Size-redshift relation for disk galaxies with mass log$(M/M_{\odot})> 10$.
Red dots show
median value in each redshift bin. Solid line shows the best-fitting size evolution (1+$z$)$^{-m}$,
where $m=0.98$. Also shown are theoretical curves if sizes evolve as 
$r\propto H(z)^{-1}$~(dashed line) and $r\propto H(z)^{-2/3}$~(dotted line).
\label{fig14}}
\end{figure}

After calculating mass, we select galaxies with log($M/M_{\odot}$)$>10.0$.
To check that we are comparing galaxy samples with similar mass, we
calculate the median mass in each redshift bin. We find masses 
log($M/M_{\odot}$)=10.5, 10,4, 10.3, and 10.3, at redshifts $z\sim0.55$, 
$z\sim1.14$, $z\sim1.75$, and $z\sim2.23$, respectively. This shows that
we are comparing samples with similar masses.
 In Figure \ref{fig14}, we show the resulting size-redshift relation 
for the sample, where red dots represent the median value in each bin.
 Fitting to the functional form $r_h \propto (1+z)^{-m}$, we find 
$m=0.98\pm{0.09}$. 

This suggests a somewhat shallower evolution in the size-redshift relation 
for a mass selected sample compared to a luminosity selected sample, even
though the difference is within errors. Comparing the $m$~parameters
between the luminosity and the mass selected samples 
indicates a decrease in stellar mass-to-light ratio at higher redshift. 

The evolution found is consistent with the simple scaling 
law $R_s\propto H(z)^{-2/3}$~(which suggests $M\sim 0.9$~over our 
redshift range). 

A shallower evolution is found in the rest-frame $V-$band by Trujillo et al.
(2005), who report $m=0.30\pm0.07$~between $z\sim 3$~and $z=0$.
Comparing with the $V-$band evolution at fixed luminosity, this also
indicates a decrease in the stellar mass-to-light ratio with redshift
(Trujillo et al. 2004, 2005).

In contrast, using disk galaxies ($n<2.5$) in rest-frame $V$-band,
Barden et al. (2005) did not find any 
significant evolution in the mass-size relation over the redshift 
range $0<z<1$. Since this is derived over a different redshift range and 
rest-frame band, it is difficult to compare results.
However, we note that a non-evolution
scenario at $1.1\lsim z \lsim 2.2$~is inconsistent with our results at 
a $\sim$4$\sigma$~level.
Therefore, further investigations are needed to determine the
redshift-size evolution, in particular, it would be desirable
to cover a larger redshift range using measurements
in a single rest-frame band.

\section{Conclusions and summary}
We use the GOODS CDF-S optical and near-IR observations to derive the 
evolution of the rest-frame 1500\AA~and 2800\AA~luminosity functions,
luminosity densities and star formation rates to $z\sim 2.2$.
Taking advantage of the high resolution $HST$~ACS imaging, we also derive
the evolution the rest-frame UV morphological properties. We find:  
\begin{itemize}

\item{There is a strong evolution in the UV characteristic magnitude
with redshift. We find a brightening in $M^*_{2800}$~by $\sim 1$~mag
between $z\sim 0.3$~and $z\sim 1.7$. At both 1500\AA~and 2800\AA, we
find no significant evolution at $z\gsim 1.7$.}

\item{The rest-frame 2800\AA~luminosity density increases by a 
factor $\sim 4$~over the redshift range $0.3<z<1.7$, while at
rest-frame 1500\AA~the increase is a factor $\sim 2$~between 
$z\sim 1.1$~and $z\sim 1.7$. The evolution flattens in both wave-lengths
at $z\gsim 1.7$.}

\item{We find that the uncorrected SFR derived from 2800\AA~luminosity
is a factor $\sim 1.7$~higher than the SFR derived from 
1500\AA~luminosity. Assuming that
the difference is due to differential dust extinction, 
we find that E(B--V)=0.20$\pm0.05$~
(using a Calzetti attenuation law and a standard Salpeter IMF) can explain the 
difference in the uncorrected SFRs.}

\item{The extinction corrected SFR is a factor $\sim 6.5$ ($\sim 3.7$) higher 
than the uncorrected SFR calculated from 1500\AA~(2800\AA) luminosity 
densities. These corrections lead to a $\sim 50\%$~higher SFR compared 
to the SFRs derived from dropout samples which assume a smaller
extinction correction (e.g., Giavalisco et al. 2004a).}

\item{The SFR we derive shows a steep increase out to $z\sim 1.5-2$. Comparing
with the local value from Wyder et al. (2005), we find that the SFR is a 
factor $\gsim 5$~higher at $z\sim 1.5$~compared to $z=0$. We note, however,
that our measurements are consistent with a flat rate at $z\gsim$1.7.}

\item{We find that the fraction of galaxies with high S\'{e}rsic index 
($n>2.5$), indicating a bulge-like morphology, increases at higher redshift. 
At $z\sim 0.5$, we find that $\sim 10\%$~of the galaxies have bulge 
morphology, while the fraction is $\sim 30\%$~at $z\sim 2.2$. At the same 
time, the mean color of the galaxies gets bluer at higher redshift. 
This suggests that an increased fraction of the star formation takes place 
in objects with bulge-like morphology at high redshift. This could be the 
sign of the formation of todays elliptical population via mergers and infall 
at redshifts $z\gsim$1. At least part of the high redshift bulge systems may, 
however, evolve into the bulges of todays population of old spirals.}

\item{Investigating the luminosity-size redshift evolution of disk galaxies 
at rest-frame 2800\AA~over the redshift range $1.1<z<2.2$,
we find that galaxies become larger at lower redshift. Our results
are consistent with trends found at both higher and lower redshift
and suggests a size evolution $r_h\propto(1+z)^{-1.1}$
between redshifts $z\sim 6$~and $z=0$. This is somewhat steeper then 
the simple scaling
law $R_s\propto H(z)^{-2/3}$, expected in the hierarchical model
for a sample selected by mass.
This suggests a stellar mass-to-light ratio that decreases at higher redshift.}

\item{We find a somewhat smaller evolution $r_h\propto (1+z)^{-0.98}$~
when selecting galaxies by mass (where mass is indirectly derived from the
galaxies rest-frame colors and absolute magnitudes). This milder 
evolution supports the change in stellar mass-to-light ratio with redshift.
Compared to measurements in the literature, we find a steeper
evolution in the size evolution for the mass selected sample. However,
investigations are made at different rest-frame bands and redshift
ranges, and are therefore difficult to directly compare.}
 \end{itemize}

\acknowledgments{
We gratefully thank the referee Alice Shapley for valuable comments
and suggestions for improving the manuscript. We also thank Christopher Conselice 
for comments on the manuscript.
Support for the GOODS $HST$~Treasury program was provided by NASA 
through grants HST-GO-09425.01-A and HST-G=-09583.01 from the Space Telescope 
Science Institute, which is operated by the Association of Universities for 
Research in Astronomy under NASA contract NAS5-26555.
Based on observations collected 
  at the European Southern Observatory, Chile (ESO programmes 168.A-0485, 
  170.A-0788, 64.O-0643, 66.A-0572, 68.A-0544, 164.O-0561, 169.A-0725,
  267.A-5729 66.A-0451, 68.A-0375 164.O-0561, 267.A-5729, 169.A-0725, 
  and 64.O-0621). M.D. acknowledge support from the Spitzer
  Legacy Science Program, provided by NASA through contract 1224666 issued
  by the Jet Propulsion Laboratory, California Institute of Technology, 
  under NASA contract 1407.
}

\clearpage
\begin{deluxetable}{lccccclcc}
\tabletypesize{\scriptsize}
\tablewidth{0pt}
\tablecaption{A summary of results}
\tablecolumns{7}
\tablehead{
\colhead{Redshift bins}& \colhead{$<z>$}  & \colhead{M$^*_{2800}$} & \colhead{$\alpha$} & \colhead{$\phi_*$}& \colhead{log($\rho_{2800}$)} & \colhead{log(SFR)} \\
\colhead{}& \colhead{}  & \colhead{} & \colhead{}  & \colhead{10$^{-4}$Mpc$^{-3}$mag$^{-1}$} & \colhead{erg s$^{-1}$Hz$^{-1}$Mpc$^{-3}$} & \colhead{M$_{\odot}$yr$^{-1}$Mpc$^{-3}$}
}
\startdata
0.29-0.37 & 0.33 &$-19.25^{+0.16}_{-0.16}$ &-&21.8$^{+2.8}_{-2.4}$ & 25.84$\pm{0.08}$& $-2.02\pm{0.08}$ \\
0.46-0.63 & 0.55 &$-19.22^{+0.28}_{-0.28}$ &$-1.39^{+0.14}_{-0.13}$ &62.3$^{+26.8}_{-21.2}$ & 26.29$\pm{0.07}$& $-1.57\pm{0.07}$ \\
0.92-1.33 & 1.14 &$-19.98^{+0.06}_{-0.06}$ &-&39.9$^{+1.6}_{-1.6}$ & 26.40$\pm{0.09}$& $-1.46\pm{0.09}$ \\
1.62-1.88 & 1.75 &$-20.37^{+0.08}_{-0.10}$ &-&40.8$^{+3.3}_{-4.1}$ & 26.56$\pm{0.10}$& $-1.29\pm{0.10}$ \\
2.08-2.37 & 2.23 &$-20.24^{+0.20}_{-0.20}$ &-&33.8$^{+13.9}_{-9.5}$ & 26.43$\pm{0.13}$& $-1.42\pm{0.13}$\\
\hline
\colhead{Redshift bins}& \colhead{$<z>$}  & \colhead{M$^*_{1500}$} & \colhead{$\alpha$} & \colhead{$\phi_*$}& \colhead{log($\rho_{1500}$)} & \colhead{log(SFR)}\\
\colhead{}& \colhead{}  & \colhead{} & \colhead{}  & \colhead{10$^{-4}$Mpc$^{-3}$mag$^{-1}$} & \colhead{erg s$^{-1}$Hz$^{-1}$Mpc$^{-3}$} & \colhead{M$_{\odot}$yr$^{-1}$Mpc$^{-3}$}\\
\hline
0.92-1.33 & 1.14 &$-19.62^{+0.06}_{-0.06}$ &-&29.6$^{+1.5}_{-1.5}$ & 26.19$\pm{0.08}$& $-1.75\pm{0.09}$ \\
1.62-1.88 & 1.75 &$-20.24^{+0.34}_{-0.29}$ &$-1.48^{+0.34}_{-0.29}$  &31.1$^{+18.3}_{-14.0}$ & 26.46$\pm{0.12}$& $-1.52\pm{0.09}$ \\
2.08-2.37 & 2.23 &$-19.87^{+0.18}_{-0.18}$ &-&33.2$^{+11.3}_{-8.0}$ & 26.34$\pm{0.09}$& $-1.60\pm{0.13}$
\enddata
\tablecomments{Results are given for Hubble constant $h=0.7$. Table columns are (1) redshift range; (2) the redshift 
dividing the volume of each bin into two equal halves; (3-5) Best fitting Schechter function 
parameters. The faint-end slope, $\alpha$~is fixed to the value derived in the second lowest redshift bins at both 2800\AA~and
1500\AA; (6) luminosity density; (7) star formation rate with no extinction correction, assuming a Salpeter IMF and an 
evolving SFR history. To get extinction corrected rates, multiply results with a factor
3.7 at 2800\AA~and a factor 6.5 at 1500\AA. See text for details.
}
\label{table1}
\end{deluxetable}

\clearpage

\clearpage

\clearpage

\clearpage

\clearpage

\clearpage

\clearpage

\clearpage

\clearpage

\clearpage

\clearpage

\clearpage

\clearpage

\clearpage

\end{document}